\input lanlmac.tex
\input proof.defs

%



%




\catcode`\@=11

\font\tenmsa=msam10

\font\sevenmsa=msam7

\font\fivemsa=msam5

\font\tenmsb=msbm10

\font\sevenmsb=msbm7

\font\fivemsb=msbm5

\newfam\msafam

\newfam\msbfam

\textfont\msafam=\tenmsa  \scriptfont\msafam=\sevenmsa

  \scriptscriptfont\msafam=\fivemsa

\textfont\msbfam=\tenmsb  \scriptfont\msbfam=\sevenmsb

  \scriptscriptfont\msbfam=\fivemsb

\def\hexnumber@#1{\ifcase#1 0\or1\or2\or3\or4\or5\or6\or7\or8\or9\or

	A\or B\or C\or D\or E\or F\fi }



\font\teneuf=eufm10

\font\seveneuf=eufm7

\font\fiveeuf=eufm5

\newfam\euffam

\textfont\euffam=\teneuf

\scriptfont\euffam=\seveneuf

\scriptscriptfont\euffam=\fiveeuf

\def\frak{\ifmmode\let\next\frak@\else

 \def\next{\Err@{Use \string\frak\space only in math mode}}\fi\next}

\def\goth{\ifmmode\let\next\frak@\else

 \def\next{\Err@{Use \string\goth\space only in math mode}}\fi\next}

\def\frak@#1{{\frak@@{#1}}}

\def\frak@@#1{\fam\euffam#1}


\edef\msa@{\hexnumber@\msafam}

\edef\msb@{\hexnumber@\msbfam}

\mathchardef\boxdot="2\msa@00

\mathchardef\boxplus="2\msa@01

\mathchardef\boxtimes="2\msa@02

\mathchardef\square="0\msa@03

\mathchardef\blacksquare="0\msa@04

\mathchardef\centerdot="2\msa@05

\mathchardef\lozenge="0\msa@06

\mathchardef\blacklozenge="0\msa@07

\mathchardef\circlearrowright="3\msa@08

\mathchardef\circlearrowleft="3\msa@09

\mathchardef\rightleftharpoons="3\msa@0A

\mathchardef\leftrightharpoons="3\msa@0B

\mathchardef\boxminus="2\msa@0C

\mathchardef\Vdash="3\msa@0D

\mathchardef\Vvdash="3\msa@0E

\mathchardef\vDash="3\msa@0F

\mathchardef\twoheadrightarrow="3\msa@10

\mathchardef\twoheadleftarrow="3\msa@11

\mathchardef\leftleftarrows="3\msa@12

\mathchardef\rightrightarrows="3\msa@13

\mathchardef\upuparrows="3\msa@14

\mathchardef\downdownarrows="3\msa@15

\mathchardef\upharpoonright="3\msa@16

\mathchardef\downharpoonright="3\msa@17

\mathchardef\upharpoonleft="3\msa@18

\mathchardef\downharpoonleft="3\msa@19

\mathchardef\rightarrowtail="3\msa@1A

\mathchardef\leftarrowtail="3\msa@1B

\mathchardef\leftrightarrows="3\msa@1C

\mathchardef\rightleftarrows="3\msa@1D

\mathchardef\Lsh="3\msa@1E

\mathchardef\Rsh="3\msa@1F

\mathchardef\rightsquigarrow="3\msa@20

\mathchardef\leftrightsquigarrow="3\msa@21

\mathchardef\looparrowleft="3\msa@22

\mathchardef\looparrowright="3\msa@23

\mathchardef\circeq="3\msa@24

\mathchardef\succsim="3\msa@25

\mathchardef\gtrsim="3\msa@26

\mathchardef\gtrapprox="3\msa@27

\mathchardef\multimap="3\msa@28

\mathchardef\therefore="3\msa@29

\mathchardef\because="3\msa@2A

\mathchardef\doteqdot="3\msa@2B

\mathchardef\triangleq="3\msa@2C

\mathchardef\precsim="3\msa@2D

\mathchardef\lesssim="3\msa@2E

\mathchardef\lessapprox="3\msa@2F

\mathchardef\eqslantless="3\msa@30

\mathchardef\eqslantgtr="3\msa@31

\mathchardef\curlyeqprec="3\msa@32

\mathchardef\curlyeqsucc="3\msa@33

\mathchardef\preccurlyeq="3\msa@34

\mathchardef\leqq="3\msa@35

\mathchardef\leqslant="3\msa@36

\mathchardef\lessgtr="3\msa@37

\mathchardef\backprime="0\msa@38

\mathchardef\risingdotseq="3\msa@3A

\mathchardef\fallingdotseq="3\msa@3B

\mathchardef\succcurlyeq="3\msa@3C

\mathchardef\geqq="3\msa@3D

\mathchardef\geqslant="3\msa@3E

\mathchardef\gtrless="3\msa@3F

\mathchardef\sqsubset="3\msa@40

\mathchardef\sqsupset="3\msa@41

\mathchardef\vartriangleright="3\msa@42

\mathchardef\vartriangleleft="3\msa@43

\mathchardef\trianglerighteq="3\msa@44

\mathchardef\trianglelefteq="3\msa@45

\mathchardef\bigstar="0\msa@46

\mathchardef\between="3\msa@47

\mathchardef\blacktriangledown="0\msa@48

\mathchardef\blacktriangleright="3\msa@49

\mathchardef\blacktriangleleft="3\msa@4A

\mathchardef\vartriangle="0\msa@4D

\mathchardef\blacktriangle="0\msa@4E

\mathchardef\triangledown="0\msa@4F

\mathchardef\eqcirc="3\msa@50

\mathchardef\lesseqgtr="3\msa@51

\mathchardef\gtreqless="3\msa@52

\mathchardef\lesseqqgtr="3\msa@53

\mathchardef\gtreqqless="3\msa@54

\mathchardef\Rrightarrow="3\msa@56

\mathchardef\Lleftarrow="3\msa@57

\mathchardef\veebar="2\msa@59

\mathchardef\barwedge="2\msa@5A

\mathchardef\doublebarwedge="2\msa@5B

\mathchardef\angle="0\msa@5C

\mathchardef\measuredangle="0\msa@5D

\mathchardef\sphericalangle="0\msa@5E

\mathchardef\varpropto="3\msa@5F

\mathchardef\smallsmile="3\msa@60

\mathchardef\smallfrown="3\msa@61

\mathchardef\Subset="3\msa@62

\mathchardef\Supset="3\msa@63

\mathchardef\Cup="2\msa@64

\mathchardef\Cap="2\msa@65

\mathchardef\curlywedge="2\msa@66

\mathchardef\curlyvee="2\msa@67

\mathchardef\leftthreetimes="2\msa@68

\mathchardef\rightthreetimes="2\msa@69

\mathchardef\subseteqq="3\msa@6A

\mathchardef\supseteqq="3\msa@6B

\mathchardef\bumpeq="3\msa@6C

\mathchardef\Bumpeq="3\msa@6D

\mathchardef\lll="3\msa@6E

\mathchardef\ggg="3\msa@6F

\mathchardef\circledS="0\msa@73

\mathchardef\pitchfork="3\msa@74

\mathchardef\dotplus="2\msa@75

\mathchardef\backsim="3\msa@76

\mathchardef\backsimeq="3\msa@77

\mathchardef\complement="0\msa@7B

\mathchardef\intercal="2\msa@7C

\mathchardef\circledcirc="2\msa@7D

\mathchardef\circledast="2\msa@7E

\mathchardef\circleddash="2\msa@7F

\def\ulcorner{\delimiter"4\msa@70\msa@70 }

\def\urcorner{\delimiter"5\msa@71\msa@71 }

\def\llcorner{\delimiter"4\msa@78\msa@78 }

\def\lrcorner{\delimiter"5\msa@79\msa@79 }

\def\yen{\mathhexbox\msa@55 }

\def\checkmark{\mathhexbox\msa@58 }

\def\circledR{\mathhexbox\msa@72 }

\def\maltese{\mathhexbox\msa@7A }

\mathchardef\lvertneqq="3\msb@00

\mathchardef\gvertneqq="3\msb@01

\mathchardef\nleq="3\msb@02

\mathchardef\ngeq="3\msb@03

\mathchardef\nless="3\msb@04

\mathchardef\ngtr="3\msb@05

\mathchardef\nprec="3\msb@06

\mathchardef\nsucc="3\msb@07

\mathchardef\lneqq="3\msb@08

\mathchardef\gneqq="3\msb@09

\mathchardef\nleqslant="3\msb@0A

\mathchardef\ngeqslant="3\msb@0B

\mathchardef\lneq="3\msb@0C

\mathchardef\gneq="3\msb@0D

\mathchardef\npreceq="3\msb@0E

\mathchardef\nsucceq="3\msb@0F

\mathchardef\precnsim="3\msb@10

\mathchardef\succnsim="3\msb@11

\mathchardef\lnsim="3\msb@12

\mathchardef\gnsim="3\msb@13

\mathchardef\nleqq="3\msb@14

\mathchardef\ngeqq="3\msb@15

\mathchardef\precneqq="3\msb@16

\mathchardef\succneqq="3\msb@17

\mathchardef\precnapprox="3\msb@18

\mathchardef\succnapprox="3\msb@19

\mathchardef\lnapprox="3\msb@1A

\mathchardef\gnapprox="3\msb@1B

\mathchardef\nsim="3\msb@1C


\mathchardef\ncong="3\msb@1D

\mathchardef\varsubsetneq="3\msb@20

\mathchardef\varsupsetneq="3\msb@21

\mathchardef\nsubseteqq="3\msb@22

\mathchardef\nsupseteqq="3\msb@23

\mathchardef\subsetneqq="3\msb@24

\mathchardef\supsetneqq="3\msb@25

\mathchardef\varsubsetneqq="3\msb@26

\mathchardef\varsupsetneqq="3\msb@27

\mathchardef\subsetneq="3\msb@28

\mathchardef\supsetneq="3\msb@29

\mathchardef\nsubseteq="3\msb@2A

\mathchardef\nsupseteq="3\msb@2B

\mathchardef\nparallel="3\msb@2C

\mathchardef\nmid="3\msb@2D

\mathchardef\nshortmid="3\msb@2E

\mathchardef\nshortparallel="3\msb@2F

\mathchardef\nvdash="3\msb@30

\mathchardef\nVdash="3\msb@31

\mathchardef\nvDash="3\msb@32

\mathchardef\nVDash="3\msb@33

\mathchardef\ntrianglerighteq="3\msb@34

\mathchardef\ntrianglelefteq="3\msb@35

\mathchardef\ntriangleleft="3\msb@36

\mathchardef\ntriangleright="3\msb@37

\mathchardef\nleftarrow="3\msb@38

\mathchardef\nrightarrow="3\msb@39

\mathchardef\nLeftarrow="3\msb@3A

\mathchardef\nRightarrow="3\msb@3B

\mathchardef\nLeftrightarrow="3\msb@3C

\mathchardef\nleftrightarrow="3\msb@3D

\mathchardef\divideontimes="2\msb@3E

\mathchardef\varnothing="0\msb@3F

\mathchardef\nexists="0\msb@40

\mathchardef\mho="0\msb@66

\mathchardef\eth="0\msb@67

\mathchardef\eqsim="3\msb@68

\mathchardef\beth="0\msb@69

\mathchardef\gimel="0\msb@6A

\mathchardef\daleth="0\msb@6B

\mathchardef\lessdot="3\msb@6C

\mathchardef\gtrdot="3\msb@6D

\mathchardef\ltimes="2\msb@6E

\mathchardef\rtimes="2\msb@6F

\mathchardef\shortmid="3\msb@70

\mathchardef\shortparallel="3\msb@71

\mathchardef\smallsetminus="2\msb@72

\mathchardef\thicksim="3\msb@73

\mathchardef\thickapprox="3\msb@74

\mathchardef\approxeq="3\msb@75

\mathchardef\succapprox="3\msb@76

\mathchardef\precapprox="3\msb@77

\mathchardef\curvearrowleft="3\msb@78

\mathchardef\curvearrowright="3\msb@79

\mathchardef\digamma="0\msb@7A

\mathchardef\varkappa="0\msb@7B

\mathchardef\hslash="0\msb@7D

\mathchardef\hbar="0\msb@7E

\mathchardef\backepsilon="3\msb@7F

\def\Bbb{\ifmmode\let\next\Bbb@\else

 \def\next{\errmessage{Use \string\Bbb\space only in math mode}}\fi\next}

\def\Bbb@#1{{\Bbb@@{#1}}}

\def\Bbb@@#1{\fam\msbfam#1}

\catcode`\@=12

\input epsf.tex
\overfullrule=0pt
\def\figbox#1#2{\epsfxsize=#1\vcenter{
\epsfbox{#2}}}
\newcount\figno
\figno=0
\def\fig#1#2#3{
\par\begingroup\parindent=0pt\leftskip=1cm\rightskip=1cm\parindent=0pt
\baselineskip=11pt
\global\advance\figno by 1
\midinsert
\epsfxsize=#3
\centerline{\epsfbox{#2}}
\vskip 12pt
{\bf Fig.\ \the\figno:} #1\par
\endinsert\endgroup\par
}
\def\figlabel#1{\xdef#1{\the\figno}%
\writedef{#1\leftbracket \the\figno}%
}
\def\omit#1{}

\def\e#1{{\rm e}^{#1}}
\def\pre#1{{\tt
#1}}

\def\Rc{{\check R}}
\def\qed{\hfill\vbox{\hrule height.4pt%
\hbox{\vrule width.4pt height3pt \kern3pt\vrule width.4pt}\hrule height.4pt}\medskip}
%
%
\def\Kr{Krattenthaler}
\lref\BIBLE{D. Bressoud, {\sl Proofs and confirmations. The story of the alternating
sign matrix conjecture}, Cambridge University Press (1999).}
\lref\KUP{G. Kuperberg, {\sl Another proof of the alternating sign matrix conjecture},
Int. Math. Research Notes (1996) 139-150.}
\lref\KUPb{G. Kuperberg, {\sl Symmetry classes of alternating-sign matrices under one roof},
{\it Ann. of Math.} (2) 156 (2002), no. 3, 835--866,
\pre{math.CO/0008184}.}
\lref\RS{A.V. Razumov and Yu.G. Stroganov, 
{\sl Combinatorial nature
of ground state vector of $O(1)$ loop model},
{\it Theor. Math. Phys.} 
{\bf 138} (2004) 333-337; {\it Teor. Mat. Fiz.} 138 (2004) 395-400, \pre{math.CO/0104216}.}
\lref\PRdG{P. A. Pearce, V. Rittenberg and J. de Gier, 
{\sl Critical Q=1 Potts Model and Temperley--Lieb Stochastic Processes},
\pre{cond-mat/0108051}.}
\lref\RSb{A.V. Razumov and Yu.G. Stroganov,
{\sl $O(1)$ loop model with different boundary conditions 
and symmetry classes of alternating-sign matrices},
\pre{cond-mat/0108103}.}
\lref\IZER{A. Izergin, {\sl Partition function of the six-vertex
model in a finite volume}, {\it Sov. Phys. Dokl.} {\bf 32} (1987) 878-879.}
\lref\KOR{V. Korepin, {\sl Calculation of norms of Bethe wave functions},
{\it Comm. Math. Phys.} {\bf 86} (1982) 391-418.}
\lref\STROIK{Yu.G. Stroganov,
{\sl A new way to deal with Izergin--Korepin determinant at root of unity},
\pre{math-ph/0204042}, and {\sl Izergin--Korepin determinant reloaded},
\pre{math-ph/0409072}.}
\lref\OKADA{S. Okada, {\sl Enumeration of Symmetry Classes of 
Alternating Sign Matrices and Characters of Classical Groups}, \pre{math.CO/0408234}.}
\lref\MNosc{S. Mitra and B. Nienhuis, {\sl 
Osculating random walks on cylinders}, in
{\it Discrete random walks}, 
DRW'03, C. Banderier and
C. Krattenthaler edrs, Discrete Mathematics and Computer Science
Proceedings AC (2003) 259-264, \pre{math-ph/0312036}.} 
\lref\MNdGB{S. Mitra, B. Nienhuis, J. de Gier and M.T. Batchelor,
{\sl Exact expressions for correlations in the ground state 
of the dense $O(1)$ loop model}, 
JSTAT (2004) P09010,
\pre{cond-math/0401245}.}
\lref\BdGN{M.T. Batchelor, J. de Gier and B. Nienhuis,
{\sl The quantum symmetric XXZ chain at $\Delta=-1/2$, alternating sign matrices and 
plane partitions},
{\it J. Phys.} A34 (2001) L265--L270,
\pre{cond-mat/0101385}.}
\lref\dG{J.~de~Gier, {\sl Loops, matchings and alternating-sign matrices},
\pre{math.CO/0211285}.}
\lref\Wie{B. Wieland, {\it  A large dihedral symmetry of the set of
alternating-sign matrices}, 
Electron. J. Combin. {\bf 7} (2000) R37, \pre{math.CO/0006234}.}
\lref\LGV{B. Lindstr\"om, {\it On the vector representations of
induced matroids}, Bull. London Math. Soc. {\bf 5} (1973)
85-90\semi
I. M. Gessel and X. Viennot, {\it Binomial determinants, paths and
hook formulae}, Adv. Math. { \bf 58} (1985) 300-321. }
\lref\DFZJZ{P.~Di~Francesco, P.~Zinn-Justin and J.-B.~Zuber,
{\sl A Bijection between classes of Fully Packed Loops and Plane Partitions},
{\it Electron. J. Combi.} to appear, \pre{math.CO/0311220}.}
\lref\twostep{J. Grassberger, A. King and P. Tirao, {\sl On the homology of
free 2-step nilpotent Lie algebras}, {\it J. Algebra} {\bf  254}, 213--225 (2002).}
{
\lref\Sloane{On-Line Encyclopedia of Integer Sequences,\hfill\break
{\tt http://www.research.att.com/$\sim$njas/sequences/Seis.html}%
.}
}
\lref\Krtr{
 M. Ciucu, C.\Kr, T. Eisenk\"olbl and D. Zare, 
{\sl Enumeration of lozenge tilings of hexagons with a central triangular hole}, 
{\it J. Combin. Theory Ser.} A 95 (2001), 251--334,
\pre{math.CO/9910053}\semi
C. \Kr, {\sl Descending plane partitions and rhombus tilings of a hexagon with triangular hole},
\pre{math.CO/0310188}.}
\lref\Krdet{C. \Kr,  {\sl Advanced determinant calculus}, 
{\it S\'eminaire Lotharingien Combin.} 42 (``The Andrews Festschrift") (1999), Article B42q,
\pre{math.CO/9902004}.}
\lref\Krcorn{M. Ciucu and C. \Kr, {\sl Enumeration of lozenge tilings of hexagons with cut-off corners}, 
{\it J. Combin. Theory} Ser. A 100 (2002), 201--231, \pre{math.CO/0104058}.}
\lref\DFZ{P.~Di~Francesco and J.-B.~Zuber, {\sl On FPL 
configurations with four sets of nested arches}, 
JSTAT (2004) P06005, \pre{cond-mat/0403268}.}
\lref\DFZJZb{P.~Di~Francesco, P.~Zinn-Justin and J.-B.~Zuber,
{\sl Determinant Formulae for some Tiling Problems and Application to Fully Packed Loops},
\pre{math-ph/0410002}.}
\lref\AvM{M.~Adler and P.~van~Moerbeke, {\sl Virasoro action on Schur function expansions, 
skew Young tableaux and random walks},
\pre{math.PR/0309202}.}
\lref\Kratt{F. Caselli and C.~\Kr, {\sl Proof of two conjectures of
Zuber on fully packed loop configurations}, {J. Combin. Theory
Ser.} {\bf A 108} (2004), 123--146, \pre{math.CO/0312217}.  }
\lref\PDFone{P. Di Francesco, {\it A refined Razumov--Stroganov conjecture}, JSTAT P006009 (2004),
\pre{cond-mat/0407477}.}
\lref\PDFtwo{P. Di Francesco, {\it A refined Razumov--Stroganov conjecture II},
\pre{cond-mat/0409576}.}
\lref\BRAU{J. De Gier and B. Nienhuis, {\sl Brauer loops and the commuting variety},
\pre{math.AG/0410392}.}
%
%
%
\Title{SPhT-T04/133}
{\vbox{
\centerline{Around the Razumov--Stroganov conjecture:}
\medskip
\centerline{proof of a multi-parameter sum rule}
}}
{\bigskip
\centerline{P.~Di~Francesco}
\smallskip
\centerline{Service de Physique Th\'eorique de Saclay,
CEA/DSM/SPhT, URA 2306 du CNRS}
\centerline{C.E.A.-Saclay, F-91191 Gif sur Yvette Cedex, France}
\smallskip
\centerline{P. Zinn-Justin}
\smallskip
\centerline{LIFR--MIIP, Independent University, 
119002, Bolshoy Vlasyevskiy Pereulok 11, Moscow, Russia}
\centerline{and Laboratoire de Physique Th\'eorique et Mod\`eles Statistiques, UMR 8626 du CNRS}
\centerline{Universit\'e Paris-Sud, B\^atiment 100,  F-91405 Orsay Cedex, France}}
\medskip

\bigskip\bigskip\bigskip
\noindent
We prove that the sum of entries of the suitably normalized
groundstate vector of the $O(1)$ loop
model with periodic boundary conditions 
on a periodic strip of size $2n$ is equal to the total number of 
$n\times n$ alternating sign matrices. This is done by identifying
the state sum of
a multi-parameter inhomogeneous version of the $O(1)$ model with the
partition function of the inhomogeneous six-vertex model on a 
$n\times n$ square grid with domain wall boundary conditions.

\bigskip\bigskip\bigskip

\font\eightrm=cmr8
\centerline{\eightrm AMS Subject Classification (2000): Primary 05A19; Secondary 52C20, 82B20}

\Date{10/2004}
%
%
\newsec{Introduction}
Alternating Sign Matrices (ASM), i.e.\ matrices
with entries $0,1,-1$, such that $1$ and $-1$'s alternate along each row and column, 
possibly separated by arbitrarily many $0$'s, and such that row and column sums are all $1$, 
have attracted much
attention over the years and seem to be a Leitmotiv of modern combinatorics,
hidden in many apparently unrelated problems, involving among others various types
of plane partitions
or the rhombus tilings of domains of the plane (see the beautiful book by Bressoud \BIBLE\
and references therein). The intrusion first of physics and then
of physicists in the subject was due to the fundamental remark that the ASM of size $n\times n$
may be identified with configurations of the six-vertex model, that consist of putting
arrows on the edges of a $n\times n$ square grid, subject to the ice rule 
(there are exactly two incoming and two outgoing
arrows at each vertex of the grid), 
with so-called domain wall boundary conditions.
This remark was instrumental
in Kuperberg's alternative proof of the ASM conjecture \KUP.  The latter relied
crucially on the integrability property of this model, that eventually allowed for finding 
closed determinantal expressions for
the total number $A_n$ of ASM of size $n\times n$, and some of its refinements.
This particular version of the six-vertex model has been extensively studied 
by physicists, culminating in a multi-parameter determinant formula for the partition
function of the model, due to Izergin and Korepin \IZER\ \KOR; some of its specializations were more
recently studied by Okada \OKADA\ and Stroganov \STROIK. An interesting alternative formulation of the model
is in terms of Fully Packed Loops (FPL). The configurations of this model are obtained by 
occupying or not the edges of the grid with bonds, with the constraint that exactly
two bonds are incident to each vertex of the grid. The model is moreover
subject to the boundary condition that every other external edge around the grid is occupied
by a bond. These are then labeled $1,2,\ldots,2n$. A given configuration realizes a pairing
of these external bonds via non-intersecting paths of consecutive bonds, possibly separated
by closed loops.

On an apparently disconnected front, Razumov and Stroganov \RS\ discovered a remarkable
combinatorial structure hidden in the groundstate vector of the
homogeneous $O(1)$ loop model, surprisingly also related to ASM numbers. The latter model
may be expressed in terms of a purely algebraic Hamiltonian, which is nothing but the sum of
generators of the Temperley--Lieb algebra, acting on the Hilbert space of link patterns $\pi$,
i.e.\ planar diagrams of $2n$ points around a circle connected by pairs via
non-intersecting arches across the disk. These express the net connectivity pattern of
the configurations of the $O(1)$ loop model on a semi-infinite cylinder of perimeter $2n$
(i.e.\ obtained by imposing periodic boundary conditions).
Razumov and Stroganov noticed that the entry of the suitably normalized groundstate vector 
$\Psi_n$ corresponding to the link pattern $\pi$ was nothing but the partition function
of the FPL model in which the external bonds are connected via the
{\it same}\/ link pattern $\pi$.  A weaker version of this conjecture, which we refer to as the 
sum rule, is that the sum of entries of $\Psi_n$
is equal to the total number $A_n$ of ASM. The sum rule was actually conjectured earlier
in \BdGN.

Both sides of this story have been generalized in various directions since the original
works. In particular, it was observed that some choices of boundary conditions in the $O(1)$ model
are connected in analogous ways to symmetry classes of ASM \refs{\PRdG,\RSb}.
Concentrating on periodic boundary conditions, it was observed recently that the 
Razumov--Stroganov conjecture could be extended by introducing anisotropies in the $O(1)$
loop model, in the form of extra bulk parameters \refs{\PDFone, \PDFtwo}.

The aim of this paper is to prove the sum rule 
conjecture of \BdGN\ in the case of periodic boundary conditions, 
and actually a generalization thereof that identifies the partition function of the
six-vertex model with domain wall boundary conditions
with the sum of entries of the groundstate vector
of a suitably defined multi-parameter inhomogeneous version of the $O(1)$ loop model.
This proves in particular the generalizations of the sum rule 
conjectured in \refs{\PDFone, \PDFtwo}. Our proof, like Kuperberg's proof of the ASM
conjecture, is non-combinatorial in nature and relies on the integrability of the model
under the form of Yang--Baxter and related equations.

The paper is organized as follows. In Sect.~2 we recall some known facts about 
the partition function $Z_n$ of the inhomogeneous 
six-vertex model with domain wall boundary conditions, 
including some simple recursion relations that
characterize it completely. In Sect.~3, we introduce the multi-parameter
inhomogeneous version of the $O(1)$ loop model and compute its transfer matrix (Sect.~3.1),
and make a few observations on the corresponding groundstate vector $\Psi_n$ (Sect.~3.2), 
in particular that the sum of entries of this vector, once suitably normalized, 
coincides with $Z_n$. This section is completed by appendix A,
where we display the explicit groundstate vector of the $O(1)$ loop model for $n=2$, $3$. 
Section 3.3 is devoted to the proof
of this statement: we first show that the entries $\Psi_{n,\pi}$ of the vector $\Psi_n$ obey some
recursion relations relating $\Psi_{n,\pi}$ to $\Psi_{n-1,\pi'}$, when two consecutive spectral parameters
take particular relative values, and where $\pi'$ is obtained
from $\pi$ by erasing a ``little arch" connecting two corresponding consecutive points. 
As eigenvectors are always defined up to multiplicative normalizations, we have to fix precisely
the relative normalizations of $\Psi_n$ and $\Psi_{n-1}$ in the process. This is done by computing
the degree of $\Psi_n$ as a homogeneous polynomial of the spectral parameters of the model,
and involves deriving an upper bound for this degree (the calculation, based on
the Algebraic Bethe Ansatz formulation of $\Psi_n$, is detailed in appendix B), 
and showing that no extra non-trivial polynomial normalization is allowed by this bound.
This is finally used
to prove that the sum of entries of $\Psi_n$ is a symmetric homogeneous polynomial
of the spectral parameters and that it obeys the {\it same}\/ recursion
relations as the six-vertex partition function $Z_n$. The sum rule follows. Further recursion
properties are briefly discussed.
Section 3.4 displays a few applications of these results, including the proof of the
conjecture on the sum of components,
and some of its recently conjectured generalizations.
A few concluding remarks are gathered in Sect.~4.

\newsec{Six Vertex model with Domain Wall Boundary Conditions}

The configurations of the six vertex (6V) model on the square lattice are obtained
by orienting each edge of the lattice with arrows, such that at each vertex
exactly two arrows point to (and two from) the vertex.
These are weighted according to the six possible vertex configurations below
$$\figbox{10.cm}{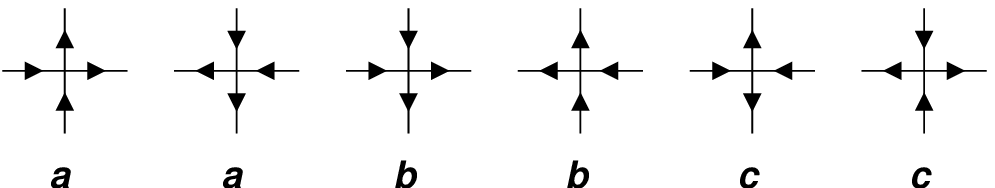}$$
with $a$, $b$, $c$ given by
\eqn\abc{
a=q^{-1/2}w-q^{1/2}z\qquad b=q^{-1/2}z-q^{1/2}w\qquad c=(q^{-1}-q)(z\,w)^{1/2}
}
and
where $w$, $z$ are the horizontal and vertical spectral parameters of the vertex. $q$ is
an additional global parameter, independent of the vertex.\foot{Note that
we use a slightly unusual sign convention for $q$, which is however convenient here.}

A case of particular interest is when the model is defined on a square $n\times n$
grid, with so-called domain wall boundary conditions (DWBC),
namely with horizontal external edges pointing
inwards and vertical external edges pointing outwards. 
Moreover, we consider the fully inhomogeneous case where we pick $n$ arbitrary horizontal
spectral parameters, one for each row say $z_1,\ldots,z_n$
and $n$ arbitrary vertical spectral parameters, one for each 
column say $z_{n+1},\ldots,z_{2n}$.

The partition function $Z_n(z_1,\ldots,z_{2n})$ of this model was computed by Izergin \IZER\
using earlier work of  Korepin \KOR\ and
takes the form of a determinant (IK determinant), which is symmetric in the sets $z_1,\ldots,z_n$
and $z_{n+1},\ldots,z_{2n}$.
It is a remarkable property, first discovered by Okada \OKADA, that when $q=\e{2 i \pi/3}$,
the partition function is actually
fully symmetric in the $2n$ horizontal and vertical
spectral parameters $z_1,z_2,\ldots,z_{2n}$.
It can be identified \refs{\STROIK, \OKADA}, up to a factor $(-1)^{n(n-1)/2}(q^{-1}-q)^n \prod_{i=1}^{2n} z_i^{1/2}$ 
which in the present work we absorb in the normalization of the partition function,
as the Schur function of the spectral parameters corresponding to
the Young diagram $Y_n$ with two rows of length $n-1$, two rows of length $n-2$, $\ldots$,
two rows of length $2$ and two rows of length $1$:
\eqn\ikpf{Z_n(z_1,\ldots,z_{2n}) = s_{Y_n}(z_1,\ldots,z_{2n})\ . }

The study of the cubic root of unity case has been extremely fruitful \refs{\KUP, \STROIK},
allowing for 
instance to find various generating functions for (refined) numbers of alternating 
sign matrices (ASM), in bijection with the 6V configurations with DWBC. In particular,
when all parameters $z_i=1$, the various vertex weights are all equal
and we recover simply the total number of such configurations
\eqn\asm{ 3^{-n(n-1)/2} Z_n(1,1,\ldots,1)=A_n=\prod_{i=0}^{n-1}{(3i+1)!\over (n+i)!}}
while by taking $z_1=(1+q\,t)/(q+t)$, $z_2=(1+q\,u)/(q+u)$, 
and all other parameters to $1$, one gets the doubly-refined ASM number generating function
\eqn\refasm{{\big(q^2(q+t)(q+u)\big)^{n-1}\over 3^{n(n-1)/2}}
Z_n\left({1+q\,t\over q+t},{1+q\,u\over q+u},1\ldots,1\right)=A_n(t,u)=
\sum_{j=1}^n t^{j-1}u^{k-1}A_{n,j,k} }
where $A_{n,j,k}$ denotes the total number of $n\times n$ ASM with a $1$ in position $j$
on the top row (counted from left to right) and $k$ on the bottom row
counted from right to left).

Many equivalent characterizations of the IK determinant are available. Here we will make use
of the recursion relations obtained in \STROIK\ for the particular case $q=\e{2i\pi/3}$,
to which we restrict ourselves from now on, namely that
\eqn\recustro{ Z_n(z_1,\ldots,z_{2n})\big\vert_{z_{i+1}=q^2\,z_i} =
\prod_{\scriptstyle j=1\atop\scriptstyle j\neq i,i+1}^{2n}
(q z_i-z_j)\, Z_{n-1}(z_1,\ldots,z_{i-1},z_{i+2},\ldots,z_{2n})\ .}
This recursion relation and the fact that $Z_n$ is a symmetric homogeneous
polynomial in its $2n$ variables with degree $\leq n-1$ in each variable
and total degree $n(n-1)$ are sufficient to completely fix $Z_n$.

\newsec{Inhomogeneous $O(1)$ loop model}
\subsec{Model and transfer matrix}
We now turn to the $O(1)$ loop model. It is defined on a semi-infinite cylinder of square lattice,
with even perimeter $2n$ whose edge centers are labelled $1,2,\ldots,2n$ 
counterclockwise. The configurations of the model are obtained by picking any of the two
possible face configurations 
$\vcenter{\hbox{\epsfbox{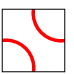}}}$ 
or 
$\vcenter{\hbox{\epsfbox{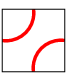}}}$ 
at
each face of the lattice. We moreover associate respective probabilities $t_i$ and $1-t_i$
to these face configurations when they sit in the $i$-th row, corresponding to the top
edge center labelled $i$.  We see that the configurations of the model form either closed
loops or open curves joining boundary points by pairs, without any intersection beteen curves.
In fact, each configuration realizes a planar pairing of the boundary points via a
link pattern, 
namely a diagram in which $2n$ labelled and regularly spaced points
of a circle are connected by pairs via non-intersecting straight segments.
Note that one does not pay attention to which way the loops wind around the cylinder,
so that the semi-infinite cylinder should really be thought of as a disk (by adding
the point at infinity).
The set of link patterns over $2n$ points is denoted by $LP_n$, and its cardinality is 
$c_n=(2n)!/(n!(n+1)!)$. We may also view $\pi\in LP_n$ as an involutive planar permutation
of the symmetric group $S_{2n}$ with only cycles of length $2$.

We may now ask what is the probability $P_n(t_1,\ldots,t_{2n}\vert \pi)$ in random configurations of 
the model that the boundary points be pair-connected according to a given link pattern 
$\pi\in LP_n$. Forming the vector 
$P_n(t_1,\ldots,t_{2n})=\{P_n(t_1,\ldots,t_{2n}\vert \pi)\}_{\pi\in LP_n}$, we immediately see that
it satisfies the eigenvector condition
\eqn\evectP{ T_n(t_1,\ldots,t_{2n}) P_n(t_1,\ldots,t_{2n})=P_n(t_1,\ldots,t_{2n}) }
where the transfer matrix $T_n$ expresses the addition of an extra row to the semi-infinite cylinder,
namely
\eqn\transmat{ T_n(t_1,\ldots,t_{2n})=\prod_{i=1}^{2n} \big(t_i 
\vcenter{\hbox{\epsfbox{mov1.eps}}}
+(1-t_i) 
\vcenter{\hbox{\epsfbox{mov2.eps}}}
\big)}
with periodic boundary conditions around the cylinder.

Let us parameterize our probabilities via
$t_i={q\,z_i-q^{-1}t\over q\,t-q^{-1}z_i}$, $1-t_i={z_i-t\over q\,t-q^{-1}z_i}$, where we recall that
$q=\e{2i\pi/3}$.
Note that for $z_i=t\,\e{-i\theta_i}$, $\theta_i\in ]0,2\pi/3[$, the weights satisfy $0<t_i<1$
and one can easily check that $T_n$ satisfies the hypotheses of the Perron--Frobenius theorem,
$P_n$ being the Perron--Frobenius eigenvector. In particular, the corresponding eigenvalue
($1$) is non-degenerate for such values of the $z_i$.
Let us also introduce the $R$-matrix
\eqn\rrcopa{ R(z,w)=
\vcenter{\hbox{\epsfxsize=1.2cm\epsfbox{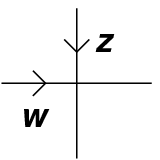}}}
=
{q\,z-q^{-1}w\over q\,w-q^{-1}z}
\vcenter{\hbox{\epsfbox{mov1.eps}}}
+{z-w\over q\,w-q^{-1}z}
\vcenter{\hbox{\epsfbox{mov2.eps}}}
\ .}
We shall often need a ``dual'' graphical depiction, in which
the $R$-matrix corresponds to the crossing of two oriented lines,
where the left (resp.\ right) incoming line carries the parameter $z$ (resp.\ $w$).
\fig{Transfer matrix as a product of $R$-matrices.%
}{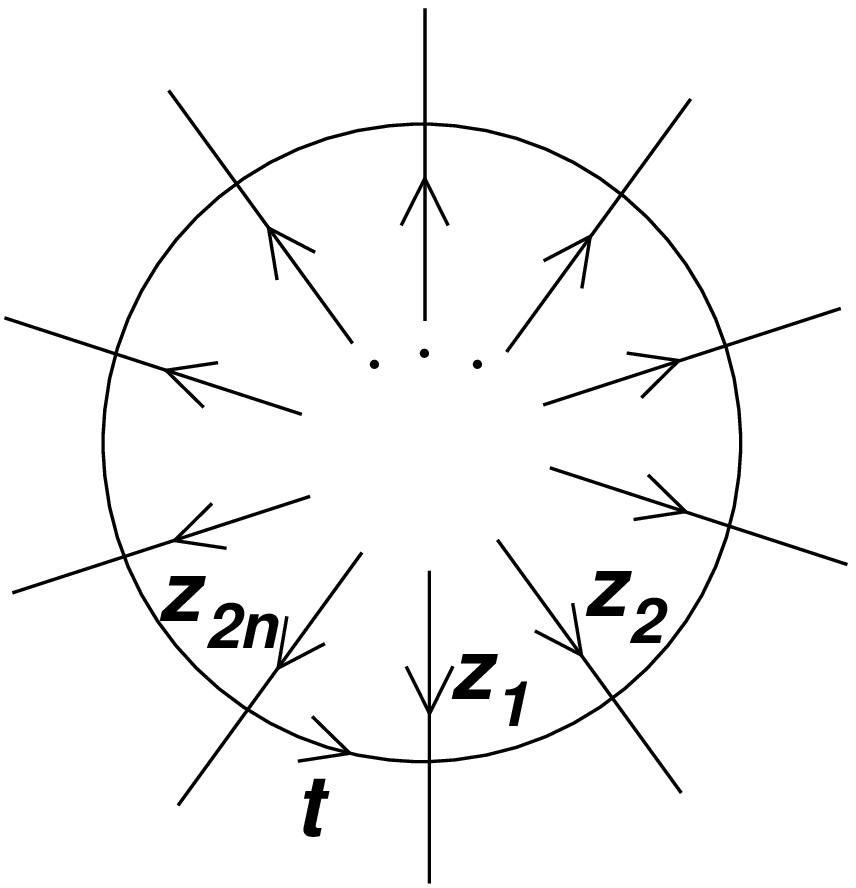}{4cm}\figlabel\flow
Then, denoting by the index $0$ an auxiliary space (propagating horizontally on the cylinder), and
$i$ the $i$-th vertical space, we can rewrite \transmat\ into
the purely symbolic expression (see Fig.~\flow)
\eqn\transmatwo{T_n\equiv T_n(t\vert z_1,\ldots,z_{2n})={\rm Tr}_0\left( R_{2n,0}(z_{2n},t)
\cdots R_{1,0}(z_1,t) \right) }
where the order of the matrices corresponds to following around the auxiliary line, 
and the trace represents closure of the auxiliary line. To avoid any possible confusion,
we note that if one ``unrolls'' the transfer matrix of Fig.~1 so that the vertices are numbered
in increasing order from left to right (with periodic boundary conditions), then the flow of time
is downwards (i.e.\ the semi-infinite cylinder is infinite in the ``up'' direction).

\subsec{Groundstate vector: empirical observations}
Solving the above eigenvector condition \evectP\ numerically (see appendix A
for the explicit values of $n=2$, $3$), we have observed the following properties.

\item{(i)} when normalized by a suitable overall multiplicative factor $\alpha_n$, the entries
of the probability vector $\Psi_n\equiv \alpha_n P_n$ 
are homogeneous polynomials in the variables $z_1,\ldots,z_{2n}$, independent of $t$, with degree $\leq n-1$
in each variable and total degree $n(n-1)$.

\item{(ii)} The factor $\alpha_n$ may be chosen so that, in addition to property (i), the sum
of entries of $\Psi_n$ be exactly equal to the partition function $Z_n(z_1,\ldots,z_{2n})$ of Sect.~2
above.

\item{(iii)} With the choice of normalization of property (ii), the entries $\Psi_{n,\pi}$ of $\Psi_n$
are such that the symmetrized sum of monomials
\eqn\monomial{ \sum_{\sigma\in S_n}\prod_{k=1}^n (z_{i_k}z_{j_k})^{\sigma(k)-1}}
where $\pi=(i_1 j_1)\cdots(i_n j_n)$,
occurs with coefficient $1$ in $\Psi_{n,\pi}$, and does not occur in 
any $\Psi_{n,\pi'}$, $\pi'\neq \pi$.

Note that the entries of $\Psi_n$ are not symmetric polynomials of the $z_i$, as opposed to their sum.
The entries $\Psi_{n,\pi}$ thus form a new family of non-symmetric polynomials, based on a monomial germ
only depending on $\pi\in LP_n$, according to the property (iii).

\fig{The transfer matrix $T$ commutes with that, $T'$, of the tilted $n$-dislocation $O(1)$ loop
model on a semi-infinite cylinder. The transfer matrix of the latter is made of
$n$ rows of tilted face operators, followed by a global rotation of one half-turn.
Each face receives the probability $t_{i,j}$ given by Eq.~\paratij\ at
the intersection of the diagonal lines $i$ and $j$, carrying the spectral parameters
$z_i$ and $z_j$ respectively as indicated. The commutation between $T$ and $T'$
(free sliding of the black horizontal
line on the blue and red ones across all of their mutual intersections) is readily
obtained by repeated application of the Yang--Baxter equation.}{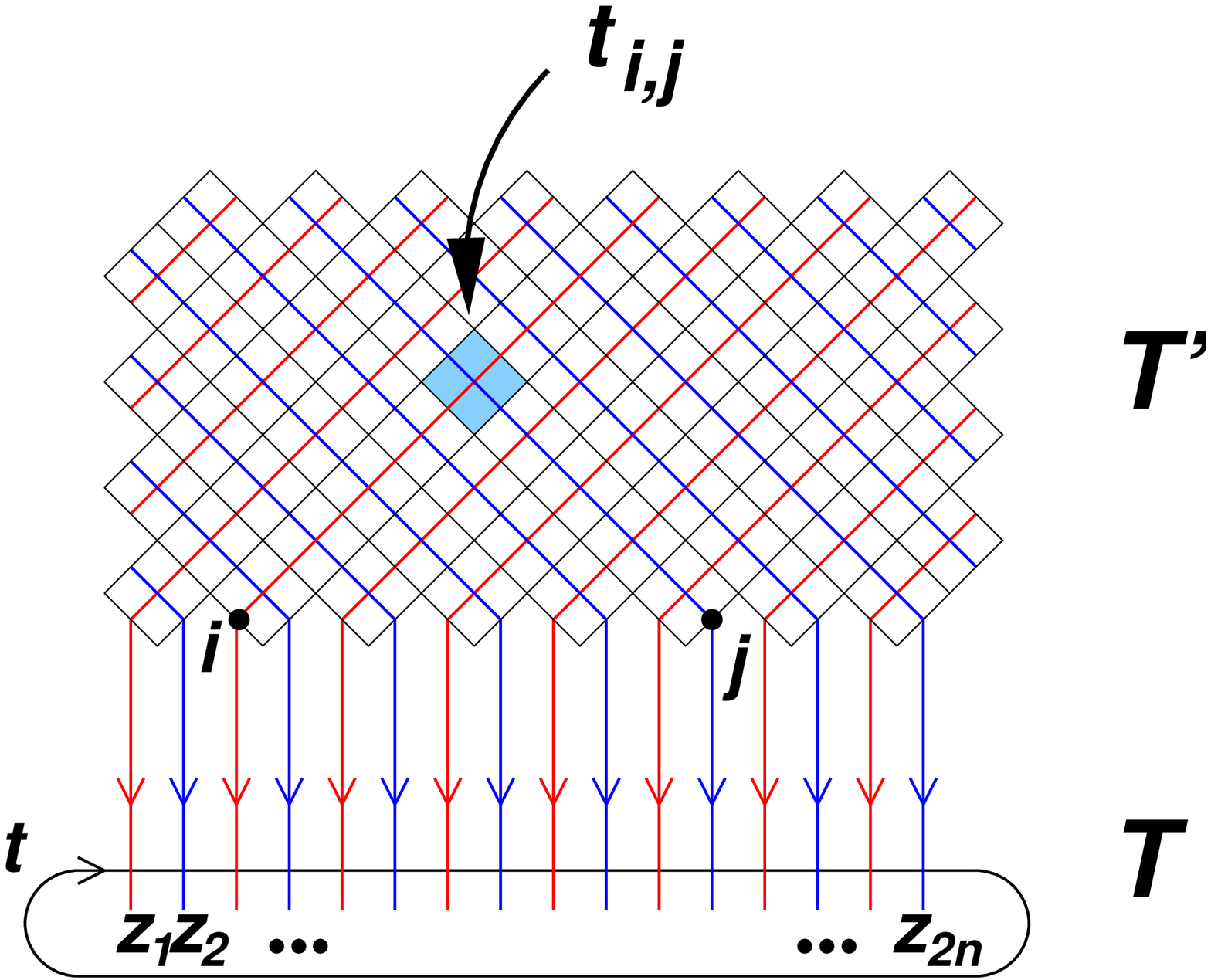}{9.cm}
\figlabel\zigzag

The fact that the entries of $\Psi_n$ do not depend on $t$ is due to the standard
property of commutation of the transfer matrices  \transmatwo\ at two distinct values of $t$, 
itself a direct consequence of the Yang--Baxter equation.
It is also possible to make the contact
between the present model and a multi-parameter version of the $O(1)$ loop model on
a semi-infinite cylinder with maximum number of dislocations introduced in \PDFtwo. 
In the latter, we simply tilt the square lattice by $45^\circ$, but keep the cylinder vertical.
This results in a zig-zag shaped boundary, with $2n$ edges still labelled $1,2,\ldots,2n$
counterclockwise, with say $1$ in the middle of an ascending edge (see Fig.\zigzag). 
The two (tilted) face configurations of the $O(1)$ loop model are still
drawn randomly with inhomogeneous probabilities $t_{i,j}$ for all the faces lying at the intersection
of the diagonal lines issued from the points $i$ ($i$ odd) and $j$ ($j$ even) of the boundary 
(these diagonal lines are wrapped around the cylinder and cross infinitely many times). If we now parametrize
\eqn\paratij{ t_{i,j}\equiv t(z_i,z_j)={q\,z_i-q^{-1}z_j\over q\,z_j-q^{-1}z_i} }
we see immediately that the transfer matrix of this model commutes with that of ours, as a direct consequence
of the Yang--Baxter equation $\vcenter{\hbox{\epsfbox{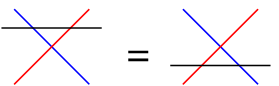}}}$. 
As no reference to $t$ is made in the latter model, we see that $\Psi_n$ must be
independent of $t$.
The tilted version of the vertex weight operator is usually understood as acting vertically
on the tensor product of left and right spaces say $i$, $i+1$, and reads
\eqn\rchech{{\check R}_{i,i+1}(z,w)=
\vcenter{\hbox{\epsfxsize=1.2cm\epsfbox{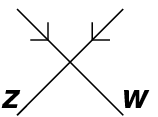}}}=
t(z,w) 
\vcenter{\hbox{\epsfbox{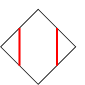}}}
+\big(1-t(z,w)\big)
\vcenter{\hbox{\epsfbox{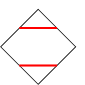}}}
=t(z,w) I +\big(1-t(z,w)\big) e_i }
where $t(z,w)$ is as in \paratij, and
$e_i$ is the Temperley--Lieb algebra generator that acts on any link pattern $\pi$
by gluing the curves that reach the points $i$ and $i+1$, and inserting a ``little arch"
that connects the points $i$ and $i+1$. Formally, one has $\check R=R{\cal P}$ where $\cal P$
is the operator that permutes the factors of the tensor product.

In the next sections, we shall set up a general framework to prove these empirical observations.

\subsec{Main properties and lemmas}

For the sake of simplicity, we rewrite the main eigenvector equation 
\evectP\ in a form manifestly polynomial in the $z_i$ and $t$, by multiplying it
by all the denominators $q\,t-q^{-1}z_i$, $i=1,2,\ldots,2n$. By a slight abuse of notation,
we still denote by $R$ and $\check R=R{\cal P}$ all the vertex weight operators in which the denominators
have been suppressed:
\eqn\rrcop{ R(z,w)=
\vcenter{\hbox{\epsfxsize=1.2cm\epsfbox{R.eps}}}
=
(q\,z-q^{-1}w)
\vcenter{\hbox{\epsfbox{mov1.eps}}}
+(z-w)
\vcenter{\hbox{\epsfbox{mov2.eps}}}
\ .}
In these notations, we now have the main equation
\eqn\main{ \left(T_n(t\vert z_1,\ldots,z_{2n})-\prod_{i=1}^{2n}(q\,t-q^{-1}z_i) I\right) \Psi_n(z_1,\ldots,z_{2n})=0}
where $T_n$ is still given by Eq.~\transmatwo\ but with $R$ as in \rrcop.
As mentioned before, for certain ranges of parameters Eq.~\main\ is
a Perron--Frobenius eigenvector equation, in which case $\Psi_n$ is uniquely defined up to normalization.
We conclude that the locus of degeneracies of the eigenvalue is of codimension
greater than zero and that $\Psi_n$ is generically well-defined. 
We may always choose the overall normalization of the eigenvector to ensure that it is a homogeneous polynomial
of all the $z_i$ (the entries $\Psi_{n,\pi}$ of $\Psi_n$ are proportional to minors of the matrix 
that annihilates $\Psi_n$, and therefore homogeneous polynomials). We may further assume that all the 
components of $\Psi_n$ are coprime, upon dividing out by their GCD. There remains an arbitrary numerical
constant in the normalization of $\Psi_n$, which will be fixed later.

Note finally that, using cyclic covariance of the problem under rotation around the cylinder,
one can easily show that
\eqn\cyclicov{ \Psi_{n,\pi}(z_1,z_2,\ldots,z_{2n-1},z_{2n})=\Psi_{n,r\pi}(z_{2n},z_1,\ldots,z_{2n-2},z_{2n-1}) }
where $r$ is the cyclic shift by one unit on the point labels of the link patterns
($r\pi(i+1)=\pi(i)+1$ with the convention that $2n+1\equiv 1$).

Our main tools will be the following three equations. First, the Yang--Baxter equation:
\eqn\ybe{
\vcenter{\hbox{
\epsfbox{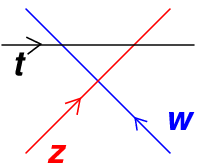}}}
=
\vcenter{\hbox{
\epsfbox{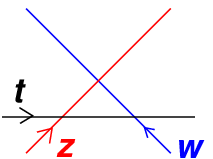}}}
}
is insensitive to the above redefinitions. 
The unitarity condition, however, is inhomogeneous:
\eqn\unit{
\vcenter{\hbox{
\epsfbox{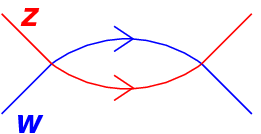}}}
=
(q\,z-q^{-1}w)(q\,w-q^{-1}z)
\vcenter{\hbox{
\epsfbox{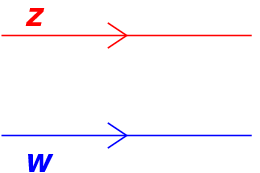}}}
}
so that ${\check R}_{i,i+1}(z,w) {\check R}_{i,i+1}(w,z)=(q\,z-q^{-1}w)(q\,w-q^{-1}z)I$.
Finally, note the crossing relation:
\eqn\inv{
\vcenter{\hbox{
\epsfbox{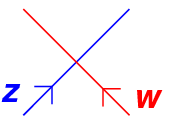}}}
=-q^{-1}
\vcenter{\hbox{
\epsfbox{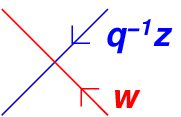}}}
}

In some figures below, orientation of lines will be omitted when it is unambiguous.
We now formulate the following fundamental lemmas:

\proclaim Lemma 1. The transfer matrices
$T_n(t|z_1,\ldots,z_i,z_{i+1},\ldots,z_{2n})$
and
$T_n(t|z_1,\ldots,z_{i+1},z_i,\ldots,z_{2n})$
are interlaced by $\check R_{i,i+1}(z_i,z_{i+1})$, namely:
\eqn\interl{T_n(t|z_1,\ldots,z_{i},z_{i+1},\ldots,z_{2n})
{\check R}_{i,i+1}(z_{i},z_{i+1})
={\check R}_{i,i+1}(z_{i},z_{i+1})
T_n(t|z_1,\ldots,z_{i+1},z_{i},\ldots,z_{2n})}

This is readily proved by a simple application of the Yang--Baxter equation:
$$\epsfbox{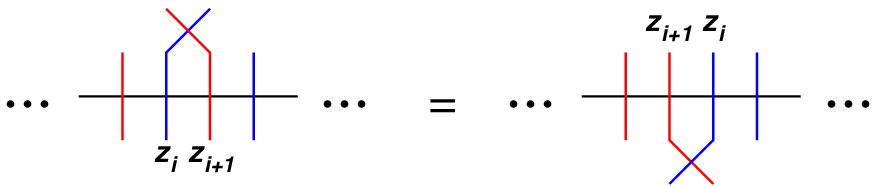}$$
\qed

To prepare the ground for recursion relations, we note that the space of link patterns
$LP_{n-1}$ is trivially embedded into $LP_n$ by simply adding a little arch say between the
points $i-1$ and $i$ in $\pi\in LP_{n-1}$, and then relabelling $j\to j+2$ the points
$j=i,i+1,\ldots,2n-2$. Let us denote by $\varphi_i$ the induced embedding of vector spaces. 
In the augmented link pattern
$\varphi_i\pi \in LP_n$, the additional little arch connects the points $i$ and $i+1$. We now have:

\proclaim Lemma 2. If two neighboring parameters $z_i$ and $z_{i+1}$ are
such that $z_{i+1}=q^2 z_i$, then
\eqn\intphi{T_n(t|z_1,\ldots,z_i,z_{i+1}=q^2 z_i,\ldots,z_{2n})\, \varphi_i
=(q^2 t-z_i)(t-z_i)\,
\varphi_i \,T_{n-1}(t|z_1,\ldots,z_{i-1},z_{i+2},z_{2n})}

The lemma is a direct consequence of unitarity and inversion relations (Eqs.~\unit--\inv). 
It is however instructive
to prove it ``by hand''.
We let the transfer matrix $T_n(t\vert z_1,\ldots,z_{2n})$
act on a link pattern $\pi\in LP_n$ with a little arch joining $i$ and $i+1$. Let us examine how
$T_n$ locally acts on this arch, namely via 
$R_{i+1,0}(q^2 z_i,t)R_{i,0}(z_i,t)$. We have
$${\raise-5mm\hbox{\epsfxsize=1.5cm\epsfbox{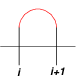}}}
=v_i u_{i+1}{\raise-3.5mm\hbox{\epsfxsize=1.5cm\epsfbox{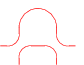}}}
+v_i v_{i+1}{\raise-3.5mm\hbox{\epsfxsize=1.5cm\epsfbox{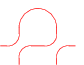}}}
+u_i u_{i+1}{\raise-3.5mm\hbox{\epsfxsize=1.5cm\epsfbox{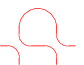}}}
+u_i v_{i+1}{\raise-3.5mm\hbox{\epsfxsize=1.5cm\epsfbox{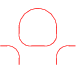}}}
$$
with $u_i=q\,z_i-q^{-1}t$ and $v_i=z_i-t$. The last three terms contribute to the same diagram,
as the loop may be safely erased (weight $1$), and the total prefactor 
$u_i u_{i+1}+v_i v_{i+1}+u_i v_{i+1}=0$
precisely at $z_{i+1}=q^2 z_i$. We are simply left with the first contribution in which the little arch
has gone across the horizontal line, while producing a factor 
$v_i u_{i+1}=(z_i-t)(q^3 z_i-q^{-1}t)=(q^2 t-z_i)(t-z_i)$
as $q^3=1$. In the process, the transfer matrix has lost the two spaces $i$ and $i+1$, 
and naturally acts on $LP_{n-1}$, while the addition of the little arch corresponds to the 
operator $\varphi_i$. 
\qed

\subsec{Recursion and factorization of the groundstate vector}
We are now ready to translate the lemmas 1 and 2 into recursion relations for the entries of $\Psi_n$.
For a given pattern $\pi$, define $E_\pi$ to be the partition of
$\{1,\ldots,2n\}$ into sequences of consecutive points not separated by little arches (see Fig.~\decomp). We
order {\it cyclically}\/ each sequence $s\in E_\pi$. 
\fig{Decomposition of a sample link pattern into sequences of consecutive points not separated
by little arches. The present example has five little arches, henceforth
five sequences $s_1=\{17,18,1\}$, $s_2=\{2,3,4,5\}$, $s_3=\{6,7,8\}$, $s_4=\{9,10,11\}$ and 
$s_5=\{12,13,14,15,16\}$.}{decomp3.eps}{5.cm}
\figlabel\decomp

\proclaim Theorem 1. The entries $\Psi_{n,\pi}$ of the groundstate
eigenvector satisfy:
\eqn\recone{\Psi_{n,\pi}(z_1,\ldots,z_{2n})=
\left(\prod_{s\in E_\pi} \prod_{\scriptstyle i,j\in s\atop
\scriptstyle i<j} (q z_i-q^{-1}z_j)\right)
\Phi_{n,\pi}(z_1,\ldots,z_{2n})}
where $\Phi_{n,\pi}$ is a polynomial which is symmetric in the set of variables
$\{ z_i, i\in s\}$ for each $s\in E_\pi$.

Let us interpret Lemma 1 by letting both sides of Eq.~\interl\
act on the groundstate vector ${\tilde \Psi}_n$, defined as $\Psi_n$ with $z_i$ and $z_{i+1}$ 
interchanged. We find that $T \Rc_{i,i+1}(z_{i},z_{i+1}){\tilde \Psi}_n=\Rc_{i,i+1}(z_{i},z_{i+1})
{\tilde T}{\tilde \Psi}_n=\Lambda \Rc_{i,i+1}(z_{i},z_{i+1}){\tilde \Psi}_n$. This shows
that $P\,\Psi_n=
 \Rc_{i,i+1}(z_{i},z_{i+1}){\tilde \Psi}_n$, where $P$,
which can only depend on $z_i$ and $z_{i+1}$,
is a polynomial because the entries of $\tilde\Psi_n$ are coprime.
To fix $P$, one can apply the transfer matrix $T'$: on the one hand we know that 
$T'\Psi_n=\prod_{i, j}(qz_{2i+1}-q^{-1}z_{2j})\Psi_n$ (we use here unnormalized $R$-matrices,
hence the extra factors);
on the other hand, applying repeatedly the identity above we find $T'\Psi_n=
\prod_{i,j} P(z_{2i+1},z_{2j})\Psi_n$. We conclude that $P(z_i,z_j)=q z_j-q^{-1}z_i$ and
\eqnn\transpo
$$\eqalignno{ (q z_{i+1}-q^{-1}z_i) &\Psi_n(z_1,\ldots,z_i,z_{i+1},\ldots,z_{2n}) \cr
&=\big((q\,z_i-q^{-1}z_{i+1})+(z_i-z_{i+1})e_i\big)\Psi_n(z_1,\ldots,z_{i+1},z_{i},\ldots,z_{2n})&\transpo}$$
In the case when $\pi$ has no little arch connecting $i$, $i+1$, we simply get
\eqn\getsimp{ (q\,z_{i+1}-q^{-1}z_i)\Psi_{n,\pi}(z_1,\ldots,z_i,z_{i+1},\ldots,z_{2n})
=(q\,z_i-q^{-1}z_{i+1})\Psi_{n,\pi}(z_1,\ldots,z_{i+1},z_{i},\ldots,z_{2n})}
hence $\Psi_{n,\pi}$ vanishes when $z_{i+1}=q^2 z_i$.

Let us now turn to the case of two points say $i$, $i+k$ within the same sequence $s$,
i.e.\ such that no little arch occurs between the points $i,i+1,\ldots,i+k$. We now use repeatedly
the Lemma 1 in order to interlace the transfer matrices at interchanged values of $z_i$ and $z_{i+k}$.
\fig{The repeated use of Yang--Baxter equation allows to
show that the operator $P_{i,k}$ intertwines $T$ at interchanged values
of $z_i$ and $z_{i+k}$. This simply amounts to letting the horizontal line slide
through all other line intersections as shown.}{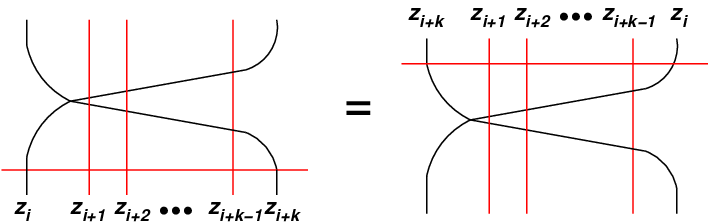}{12.cm}
\figlabel\multir
Let
\eqn\defpik{\eqalign{
&P_{i,k}(z_i,z_{i+1},\ldots,z_{i+k})
=\Rc_{i+k-1,i+k}(z_{i+k-1},z_{i+k})\Rc_{i+k-2,i+k-1}(z_{i+k-2},z_{i+k})\cdots \cr
&\cdots\Rc_{i+1,i+2}(z_{i+1},z_{i+k})\times \Rc_{i,i+1}(z_{i},z_{i+k}) 
\Rc_{i+1,i+2}(z_{i},z_{i+1}) \cdots \Rc_{i+k-1,i+k}(z_{i},z_{i+k-1})\cr}}
Then we have
\eqnn\repeatt
$$\eqalignno{
T_n(z_1,\ldots,z_{i},&\ldots,z_{i+k},\ldots,z_{2n})
P_{i,k}(z_i,\ldots,z_{i+k})\cr
&=P_{i,k}(z_i,\ldots,z_{i+k})T_n(z_1,\ldots,z_{i+k},\ldots,z_{i},\ldots,z_{2n})&\repeatt}
$$
following from the straightforward pictorial representation of Fig.~\multir.
We conclude as before that
$P{\tilde \Psi}_n$ is proportional to $\Psi_n$.
We deduce that $\Psi_n$ lies in the image of the operator $P$. But expanding
$P_{i,k}$ of Eq.~\defpik\ as a sum of products of $e$'s and $I$'s with polynomial coefficients
of the $z_i$, we find that because one of the $\Rc$ terms,
namely $\check R_{i,i+1}(z_i,z_{i+k}=q\,z_i)$,
is proportional to $e_i$,
all the link patterns contributing to the image of $P_{i,k}$ have {\it at least}\/ one 
little arch in between the points $i$ and $i+k$ (either at the first
place $j\leq i+k$, $j>i$, where a term $e_j$ is picked in the above expansion, 
or at the place $i$,
with $e_i$, if only terms $I$ have been picked before). As we have assumed $\pi$
has no such little arch in between $i$ and $i+k$, the entry of $\Psi_{n,\pi}$ must vanish, 
and this completes the proof
that $\Psi_{n,\pi}$ factors out a term $(q\, z_i-q^{-1}z_{i+k})$ when there is no little
arch in between $i$ and $i+k$ in $\pi$.

Let us now factor out all the $qz_i - q^{-1}z_j$ coming from the cancellations we have found: 
we are left with a polynomial
$\Phi_{n,\pi}$ of the $z_i$ as in Eq.~\recone\ which is, according to Eq.~\getsimp, invariant
under the interchange of $z_i$ and $z_{i+1}$. This shows that $\Phi_{n,\pi}$ of Eq.~\recone\
is symmetric under the interchange of any consecutive parameters within the same
sequence $s$, henceforth is fully symmetric in the corresponding variables. 
\qed

As a first illustration of Theorem 1, we find that in the case $\pi=\pi_0$ of the
``fully nested" link pattern that connects the points $i\leftrightarrow 2n+1-i$,
we obtain the maximal number $2{n\choose 2}=n(n-1)$ of factors from Eq.~\recone.
Up to a yet unknown polynomial $\Omega_{n,\pi_0}$ symmetric in both sets
of variables $\{z_1,\ldots,z_n\}$ and $\{z_{n+1},\ldots,z_{2n}\}$, we may write
\eqn\psipizero{ \Psi_{n,\pi_0}(z_1,\ldots,z_{2n})=\Omega_{n,\pi_0}(z_1,\ldots,z_{2n})
\prod_{1\leq i<j\leq n}
(q z_i-q^{-1}z_j) \times \prod_{n+1\leq i<j\leq 2n}(q^{-1}z_j-q\,z_i)}
where the numerical normalization factor is picked in such a way that property (iii)
of Sect.~3.2 would simply imply that $\Omega_{n,\pi_0}=1$. This will be proved below,
but for the time being the normalization of $\Psi_{n,\pi_0}$ fixes that of $\Psi_n$. 
The formula \psipizero\ extends trivially to the $n$ images 
of $\pi_0$ under rotations, $r^\ell\pi_0$, $\ell=0,1,\ldots,n-1$, by use of Eq.~\cyclicov.
Note that $r^\ell\pi_0$ has exactly two little arches joining respectively
$2n-\ell$, $2n-\ell+1$, and $n-\ell$, $n-\ell+1$.

An interesting consequence of Eq.~\transpo\ is the following:

\proclaim Theorem 2. The sum over all components of $\Psi_n$ is a symmetric polynomial
in all variables $z_1,\ldots,z_{2n}$. 

This is proved by writing Eq.~\transpo\ in components and summing over them.
We immediately get
\eqnn\compopsi
$$\eqalignno{ (q\,z_{i+1}-q^{-1}z_i) \Psi_{n,\pi}(z_1,&\ldots,z_i,z_{i+1},\ldots,z_{2n})=
(q\,z_i-q^{-1}z_{i+1})\Psi_{n,\pi}(z_1,\ldots,z_{i+1},z_{i},\ldots,z_{2n})\cr
&+(z_i-z_{i+1})\sum_{\pi'\in LP_n\atop
e_i\pi'=\pi} \Psi_{n,\pi'}(z_1,\ldots,z_{i+1},z_{i},\ldots,z_{2n})&\compopsi}$$
We now sum over all $\pi\in LP_n$, and notice that the double sum in the last term
amounts to just summing over all $\pi'\in LP_n$, without any further restriction.
Denoting by $W_n(z_1,\ldots,z_{2n})=\sum_{\pi\in LP_n}\Psi_{n,\pi}(z_1,\ldots,z_{2n})$,
we simply get that $W_n(z_1,\ldots,z_{i+1},z_i,\ldots,z_{2n})
=W_n(z_1,\ldots,z_{i},z_{i+1},\ldots,z_{2n})$. This shows the desired symmetry property,
as the full symmetric group action is generated by transpositions of neighbors.
\qed

This brings us to the main theorem of this paper, establishing recursion relations between
the entries of the groundstate vectors at different sizes $n$ and $n-1$. We have:

\proclaim Theorem 3. If two neighboring parameters $z_i$ and $z_{i+1}$ are
such that $z_{i+1}=q^2 z_i$, then either of the two following situations occur for the components
$\Psi_{n,\pi}$:\hfill\break
(i) the pattern $\pi$ has no arch joining $i$ to $i+1$,
in which case 
\eqn\vanish{\Psi_{n,\pi}(z_1,\ldots,z_i,z_{i+1}=q^2 z_i,\ldots,z_{2n})=0\ ;}
(ii) the pattern $\pi$ has a little arch joining $i$ to $i+1$, 
in which case
\eqnn\recure
$$\eqalignno{\Psi_{n,\pi}(z_1,&\ldots,z_i,z_{i+1}=q^2 z_i,\ldots,z_{2n})=\cr
&
\left(\prod_{\scriptstyle k=1\atop\scriptstyle k\ne i,i+1}^{2n} (q\,z_i-z_k)\right)
\ \Psi_{n-1,\pi'}(z_1,\ldots,z_{i-1},z_{i+2},\ldots,z_{2n})&\recure}$$
where $\pi'$ is the link pattern $\pi$ with the little arch $i$, $i+1$
removed ($\pi=\varphi_i\pi'$, $\pi'\in LP_{n-1}$).

Note that Eq.~\recure\ fixes recursively the numerical constant in the normalization of $\Psi_n$, starting from
$\Psi_1\equiv 1$.
The situation (i) is already covered by Theorem 1 above. To study the situation (ii),
we use the Lemma 2 above, and let both sides of Eq.~\intphi\ act on 
$\Psi_{n-1}\equiv\Psi_{n-1}(z_1,\ldots,z_{i-1},z_{i+2},\ldots,z_{2n})$, groundstate vector
of $\tilde T\equiv T_{n-1}(t\vert z_1,\ldots,z_{i-1},z_{i+2},\ldots,z_{2n})$. This gives
$T\varphi_i\Psi_{n-1}=(q^2 t-z_i)(t-z_i)\varphi_i \tilde T \Psi_{n-1}=(q^2 t-z_i)(t-z_i)\tilde\Lambda\varphi_i\Psi_{n-1}=
\Lambda \varphi_i\Psi_{n-1}$, where $\tilde\Lambda=\Lambda/((q^2 t-z_i)(t-q z_{i+1}))$.
Note that $T$ is evaluated at $z_{i+1}=q^2 z_i$, in which case it leaves invariant the subspace of link
patterns with a little arch joining $i$, $i+1$. The groundstate vector $\Psi_n$ then becomes proportional
to $\varphi_i\Psi_{n-1}$, with a global proportionality factor $\beta_{n,i}$, i.e.\ $\Psi_n=\beta_{n,i}
\varphi_i\Psi_{n-1}$. The overall factors $\beta_{n,i}$ are further fixed by looking at the 
component $\Psi_{n,\pi_\ell}$ of $\Psi_n$,
with link pattern $\pi_\ell=r^\ell\pi_0$, having a little arch between $i$, $i+1$.
This corresponds to taking for instance $\ell=n-i$.
We find that $\beta_{n,i}=\prod_{k\ne i, i+1}(q\, z_i-z_k)
\Omega_{n,\pi_{n-i}}\vert_{z_{i+1}=q\,z_i}/\Omega_{n-1,\pi_{n-i}'}$,
with $\pi_\ell=\varphi_{n-\ell} \pi_\ell'$.
After possibly reducing the fraction 
$\Omega_{n,\pi_{n-i}}\vert_{z_{i+1}=q\,z_i}/\Omega_{n-1,\pi_{n-i}'}=U_{n,i}/V_{n,i}$
(where both $U_{n,i}$ and $V_{n,i}$ are polynomial)
we get that
$\Psi_n/U_{n,i}=\prod_{k\ne i, i+1}(q\, z_i-z_k)\varphi_i\Psi_{n-1}/V_{n,i}$
is a polynomial, hence the poles introduced by dividing out $U_{n,i},V_{n,i}$
must be canceled by zeros of $\Psi_n$ and $\varphi_i\Psi_{n-1}$ respectively, which shows that 
$V_{n,i}$, a polynomial of $z_1,\ldots,z_{i-1},z_{i+2},\ldots,z_{2n}$, must divide $\Psi_{n-1}$, hence is
a constant, by our assumption that the entries of $\Psi_{n-1}$ are coprime. Absorbing
it into a redefinition of $U_{n,i}$, we get 
$\Omega_{n,\pi_{n-i}}\vert_{z_{i+1}=q^2 z_i}=U_{n,i}\Omega_{n-1,\pi_{n-i}'}$, for some polynomial 
$U_{n,i}\equiv U_{n}(z_1,\ldots,z_{i-1},z_{i+2},\ldots,z_n\vert z_i)$,
and the recursion relation for $z_{i+1}=q^2 z_i$ reads
\eqn\recumoins{ \Psi_{n,\pi}=U_{n,i} 
\prod_{\scriptstyle k=1\atop\scriptstyle k\ne i,i+1}^{2n} (q\, z_i-z_k) \Psi_{n-1,\pi'}\ .}

We will now proceed and show that all polynomials $U_{n,i}=1$.
To do so, we write the recursion relation \recumoins\ in the particular case of 
$\pi=\pi_n$ made of $n$ consecutive little arches
joining points $2i-1$ to $2i$, $i=1,2,\ldots,n$. Moreover, we pick the
particular values $z_{2i}=q\,z_{2i-1}$, $i=1,2,\ldots,n$ of the $z_i$. 
These allow for using Eq.~\recumoins\ iteratively $n$ times, stripping each time 
the link pattern $\pi$ from one little arch, until it is reduced to naught. 
But we may do so in 
any of $n!$ ways, according to the order in which we remove little arches from
$\pi$.
For simplicity, we set $w_i=z_{2i-1}$ from now on.
Upon removal of the $k$-th little arch, we have
\eqnn\recpit
$$\eqalignno{ &\Psi_{\pi_n}(w_1,q^2 w_1,w_2,q^2 w_2,\ldots,w_n,q^2 w_n)=
U_n(w_1,w_2,\ldots,w_{k-1},w_{k+1},\ldots,w_n\vert w_k) \times\cr
&\hskip-2.2cm
 \Big(\prod_{i=1\atop i\neq k}^n(q^2w_i-w_k)(w_i-q^2w_k) \Big) 
\Psi_{\pi_{n-1}}(w_1,q^2w_1,\ldots,w_{k-1},q^2w_{k-1},w_{k+1},q^2w_{k+1},\ldots,w_n,q^2w_n)&\recpit}$$
The $U_i$ satisfy all sorts of crossing relations, obtained by expressing removals 
of little arches in different orders. 
We adopt the notation ${\hat w}$ to express that the argument $w$ is missing
from an expression. For instance $U_n(w_1,\ldots,{\hat w}_k,\ldots,w_n\vert w_k)$ stands for the above polynomial
$U_n$ in which the argument $w_k$ is omitted from the list of $w_i$ in its first $n-1$ arguments. 
Now removing for instance the $k$-th and $m$-th little arches from $\pi$ in either order
yields the relation
\eqnn\ralkm
$$\eqalignno{
U_n(w_1,&\ldots,{\hat w}_k,\ldots,w_n\vert w_k) 
U_{n-1}(w_1,\ldots,{\hat w}_k,\ldots,{\hat w}_m\,\ldots,w_n\vert w_m)\cr
&=U_n(w_1,\ldots,{\hat w}_m,\ldots,w_n\vert w_m)
U_{n-1}(w_1,\ldots,{\hat w}_k,\ldots,{\hat w}_m\,\ldots,w_n\vert w_k)&\ralkm}$$
for all $k<m$. We shall now use these relations to prove the following

\proclaim Lemma 3. There exists a sequence of symmetric
polynomials $\alpha_j(x_1,\ldots,x_j)$, $j=1,2,\ldots,n$, such that
\eqn\Uofalpha{ U_n(w_1,\ldots,w_{n-1}\vert w_n)=\prod_{k=0}^{n-1}
\prod_{1\leq i_1<i_2<\cdots<i_k\leq n-1} \alpha_{k+1}(w_{i_1},w_{i_2},\ldots,w_{i_k},w_n) }
where, by convention, the $k=0$ term simply reads $\alpha_1(w_n)$. The other $U_n$
involved say in Eq.~\recpit\ are simply obtained by the cyclic substitution
$w_j\to w_{j+k}$ (with $w_{i+n}\equiv w_i$ for all $i$).

We will prove the lemma by induction. 
Let us however first show how to get \Uofalpha\ in the cases $n=1,2,3$.
For $n=1$, we simply define $\alpha_1(w_1)=U_1(w_1)$. For $n=2$, there are two ways of
stripping $\pi=\vcenter{\hbox{\epsfxsize=1.2cm\epsfbox{arch2-2.eps}}}$ of its two arches, yielding 
\eqn\twoways{ U_2(w_1\vert w_2)\alpha_1(w_1)=U_2(w_2\vert w_1)\alpha_1(w_2)}
therefore there exists a polynomial $\alpha_2(w_1,w_2)$, such that 
$U_2(w_1\vert w_2)=\alpha_2(w_1,w_2)\alpha_1(w_2)$ and 
$U_2(w_2\vert w_1)=\alpha_2(w_1,w_2)\alpha_1(w_1)$, which also immediately shows that 
$\alpha_2(w_1,w_2)=\alpha_2(w_2,w_1)$.
For $n=3$, we compare the various ways of stripping 
$\pi=\vcenter{\hbox{\epsfxsize=1.2cm\epsfbox{arch3-5.eps}}}$ from its three
arches, resulting in:
\eqnn\sixways
$$\eqalignno{ U_3(w_1,&w_2\vert w_3)\alpha_2(w_1,w_2)\alpha_1(w_2)\cr
&=U_3(w_1,w_3\vert w_2)
\alpha_2(w_1,w_3)\alpha_1(w_3)=U_3(w_2,w_3\vert w_1)\alpha_2(w_2,w_3)\alpha_1(w_3)&\sixways}$$
We see that both polynomials $B_{1,3}=\alpha_1(w_3)\alpha_2(w_1,w_3)$ and 
$B_{2,3}=\alpha_1(w_3)\alpha_2(w_2,w_3)$ 
divide $U_3(w_1,w_2\vert w_3)$, as they are prime with $B_{1,2}=\alpha_2(w_1,w_2)\alpha_1(w_2)$ 
(the latter does not depend on $w_3$). The least common multiple of 
$B_{1,3}$ and $B_{2,3}$ reads $LCM(B_{1,3},B_{2,3})=\alpha_2(w_1,w_3)\alpha_2(w_2,w_3)\alpha_1(w_3)$;
it is a divisor of $U_3(w_1,w_2\vert w_3)$, which must therefore be expressed as
$U_3(w_1,w_2\vert w_3)=\alpha_3(w_1,w_2,w_3)\alpha_2(w_1,w_3)\alpha_2(w_2,w_3)\alpha_1(w_3)$ 
for some polynomial $\alpha_3$.
Finally, substituting this and its cyclically rotated versions into \sixways,
we find that $\alpha_3(w_1,w_2,w_3)=\alpha_3(w_1,w_3,w_2)=\alpha_3(w_2,w_3,w_1)$, hence
$\alpha_3$ is symmetric.

Let us now turn to the general proof.
Assume \Uofalpha\ holds up to order $n-1$.
Picking for instance $1\leq k\leq n-1$ and $m=n$, Eq.~\ralkm\ implies that 
$U_n(w_1,\ldots,w_{n-1}\vert w_n)$
$U_{n-1}(w_1,\ldots{\hat w}_k\ldots,w_{n-1}\vert w_k)
=U_n(w_1,\ldots{\hat w}_k\ldots,w_n\vert w_k)
U_{n-1}(w_1,\ldots{\hat w}_k\ldots,w_{n-1}\vert w_n)$.
The main fact here is that the polynomials $A_{n,k}\equiv U_{n-1}(w_1,\ldots{\hat w}_k\ldots,w_{n-1}\vert w_k)$
and $B_{n,k}\equiv U_{n-1}(w_1,\ldots{\hat w}_k\ldots,w_{n-1}\vert w_n)$, both expressed via \Uofalpha\
at order $n-1$ in terms of products of symmetric polynomials are actually coprime.
Indeed, $B_{n,k}$ depends explicitly on $w_n$ (and does so symmetrically within each of its 
$\alpha_j$ factors), while $A_{n,k}$ does not. We deduce that $B_{n,k}$ must divide  
$U_n(w_1,\ldots,w_{n-1}\vert w_n)$, and this is true for all $k=1,2,\ldots,n-1$,
henceforth also for their least common multiple:
\eqn\lcmfound{ LCM(\{B_{n,k}\}_{1\leq k\leq n-1})=\prod_{k=0}^{n-2}\prod_{1\leq i_1<i_2<\cdots<i_k\leq n-1} 
\alpha_{k+1}(w_{i_1},w_{i_2},\ldots,w_{i_k},w_n) }
obtained by applying the recursion hypothesis to all the $B_{n,k}$, $k=1,2,\ldots,n-1$. 
Therefore
there exists a polynomial $\alpha_n(w_1,w_2,\ldots,w_n)$ such that
$U_n(w_1,\ldots,w_{n-1}\vert w_n)=\alpha_n(w_1,\ldots,w_n) LCM(\{B_{n,k}\}_{1\leq k\leq n-1})$, which, together
with \lcmfound\ amounts to \Uofalpha. 
The analogous expressions for the $U_n$'s appearing in Eq.~\recpit\ are obviously
obtained by cyclically shifting the indices $w_j\to w_{j+k}$ for all $j$.
Let us finally show that $\alpha_n$ is symmetric in its
$n$ arguments. For this, let us pick another polynomial $U_n$ occurring in the recursion
relation \recpit, say upon removal of the $k$-th little arch, 
namely $U_n(w_1,\ldots{\hat w}_k\ldots,w_n\vert w_{k})$,
and express it analogously as a product of $\alpha_i$. We find
\eqnn\otherU
$$\eqalignno{
 U_n(w_1,\ldots{\hat w}_k\ldots,w_n\vert w_{k})&=\alpha_n(w_1,\ldots{\hat w}_k\ldots,w_n,w_{k})\cr
&\times\prod_{m=0}^{n-1}\prod_{\scriptstyle 1\leq i_1<\cdots<i_m\leq n
\atop\scriptstyle i_j\neq n-1, \ {\rm for}\ {\rm all}\ j} 
\alpha_{m+1}(w_{i_1},w_{i_2},\ldots,w_{i_m},w_{k})
&\otherU\cr}$$
Comparing the $U$'s obtained by removing first the arch $n$, then the arch $k$ and vice versa
leads to Eq.~\ralkm\ with $m=n$. Substituting \otherU\ and \Uofalpha\ into this relation,
we see that all the (symmetric)
$\alpha_j$ factors, $j=1,2,\ldots, n-1$, cancel out, and we are finally left with just
$\alpha_n(w_1,\ldots{\hat w}_k\ldots,w_{n},w_k)=\alpha_n(w_1,\ldots,w_{n})$. For $k=n-1$ this gives
the invariance of $\alpha_n$ under the interchange of its last two arguments. We may now
repeat the whole process with the removal of pairs of arches with numbers
$(n-2,n-1)$, $(n-3,n-2)$, \dots, $(1,2)$.
This yields the invariance of $\alpha_n$ under the interchange of any two of its consecutive
arguments, henceforth $\alpha_n$ is fully symmetric in its $n$ arguments.
\qed

Let us now denote by $a_j$ the total degree of the polynomial $\alpha_j$, then by Lemma 3
the total degree $d_n$ of $U_n$ reads
\eqn\degreeU{ d_n=\sum_{k=0}^{n-1} {n-1\choose k} a_{k+1} }
while the total degree $\delta_n$ of $\Omega_n$ is $\delta_n=\sum_{i=1}^n d_i$.
By direct computation, we have obtained the vector $\Psi_n$ explicitly for $n=2,3$
(see appendix A). These display $\Omega_n=1$, for $n\leq 3$, hence all corresponding $U_n$'s
and $\alpha_j$'s are trivial, all with value $1$.
Assuming there exists at least one non-trivial polynomial $\alpha_j$, 
then its degree is $a_j\geq 1$, with $j\geq 4$. By \degreeU, we see that
$d_n\geq {n-1\choose j-1} a_{j+1}$ for all $n\geq j$. This lower bound on the degree
$d_n$ is a polynomial
of $n$ with degree $j-1\geq 3$. In appendix B, we show that the entries of $\Psi_n$
have a degree bounded by $2n^2(n+1)$, hence are polynomials with degree at most cubic in $n$.
This contradicts the lower bound on $d_n$ that we have just obtained, as 
$\deg(\Psi_n)=n(n-1)+\deg(\Omega_n)=n(n-1)+\sum_{i=1}^n d_i$ grows at least like $n^j$,
$j\geq 4$. We conclude that no polynomial $\alpha_j$ may be non-trivial, therefore
all $\alpha_i$, $U_i$ and $\Omega_i$ are constants, which we fix to be $1$. This completes
the proof of \recure.
\qed

Note that this fixes in turn 
the normalization of $\Psi_{n,\pi_0}$ to
be $3^{n(n-1)/2}$ when all the parameters $z_j=1$, which is simply
a numerical constant compared to the normalization $1$ picked
in earlier papers \refs{\BdGN,\RS}. Futhermore, we deduce:

\proclaim Theorem 4. The components of $\Psi_n$ are homogeneous
polynomials of total degree $n(n-1)$, and of partial degree at most $n-1$ in each variable $z_i$.

The total degree has already been proved, since all components are homogeneous of the same degree and
$\Psi_{n,\pi_0}$ has been written out explicitly; and since no component is identically zero,
due to the Perron--Frobenius property for some values of the $z_i$.
We still have to show the degree $n-1$ in each variable. To do so,
let us denote by 
$\delta_n$ the maximum degree of $\Psi_n$ in each variable (it is the same for all variables
by cyclic covariance). Let moreover $s$ denote the reflection on link patterns that interchanges
$i\leftrightarrow 2n+1-i$. Reflecting the picture of our semi-infinite cylinder
simply amounts to this relabeling of points, and also to a reversal of all orientations
of lines in the various operators involved, such as the transfer matrix. This in
turn amounts in each R matrix to the interchange of parameters $(z_i,t)\to (t,z_i)$, also equivalent
up to an overall factor to $(z_i,t)\to (1/z_i,1/t)$. We therefore deduce a relation
\eqn\relainv{ \prod_{i=1}^{2n} z_i^{\delta_n} 
\Psi_{n,s\pi}\left({1\over z_{2n}},{1\over z_{2n-1}},\ldots,{1\over z_1}\right)=A_n(z_1,\ldots,z_{2n})
\Psi_{n,\pi}(z_1,\ldots,z_{2n})}
where $A_n$ is a rational fraction, independent of $\pi$, to be determined. 
As the l.h.s. of \relainv\ is a polynomial,
any denominator of $A_n$ should divide all $\Psi_{n,\pi}$ on the r.h.s., which contradicts our
hypothesis of coprimarity of components, hence $A_n$ is a polynomial. Moreover, iterating \relainv\
twice and noting that $s^2=1$, we get the inversion relation
\eqn\invpol{ A_n(z_1,\ldots,z_{2n})A_n\left({1\over z_{2n}},{1\over z_{2n-1}},\ldots,{1\over z_1}\right)=1
\ .}
Note that summing \relainv\ over $\pi\in LP_n$ yields 
\eqn\symrela{ \prod_{i=1}^{2n} z_i^{\delta_n} W_n\left({1\over z_1},\ldots,{1\over z_{2n}}\right)=
A_n(z_1,\ldots,z_{2n})W_n(z_1,\ldots,z_{2n})}
which implies that $A_n$ is a symmetric polynomial. The only symmetric polynomials that solve
\invpol\ are of the form $A_n(z_1,\ldots,z_{2n})=(z_1z_2\ldots z_{2n})^{m}$, but we immediately see
that $m=0$ from \relainv\ by definition of $\delta_n$ as the degree in each variable.
Finally, we may now equate the total degrees of both sides of \relainv, with the result
$2n\delta_n -n(n-1)=n(n-1)$, hence $\delta_n=n-1$.
\qed

We may now combine the two possibilities (i) and (ii) of Theorem 3 into properties of the sum
over components $W_n(z_1,\ldots,z_{2n})=\sum_{\pi\in LP_n}\Psi_{n,\pi}(z_1,\ldots,z_{2n})$. 
This gives the

\proclaim Theorem 5. The sum of components of $\Psi_n$ is equal to the partition function
of the six-vertex model with domain wall boundary conditions:
\eqn\thmfour{
W_n(z_1,\ldots,z_{2n})=Z_n(z_1,\ldots,z_{2n})
\ .}

The proof consists of summing over all link patterns $\pi$ the equations \recure\ and \vanish,
according to whether $\pi$ has a little arch $i$, $i+1$ or not, and noticing that the resulting 
recursion relation is equivalent to Eq.~\recustro, satisfied by the IK determinant. 
As it is moreover
symmetric and has the same degree as a polynomial of the $z_i$, we conclude that the two
are proportional, up to a numerical factor independent of $n$. 
The proportionality
factor between $W_n$ and $Z_n$ is fixed by comparing $W_1(z_1,z_2)=1$ to $Z_1(z_1,z_2)=1$ as well.
\qed

Finally, let us briefly describe general recursion relations.
So far we have only discussed recursion when two neighboring spectral parameters
$z_i$ and $z_{i+1}$ are related by $z_{i+1}=q^2 z_i$. What happens when $z_j=q^2 z_i$ for
arbitrary locations $i$ and $j$? Of course, it does not make any difference for the sum of components
since it is a symmetric function of all parameters. The components themselves, however, are not symmetric.
But lemma 1 allows us, as we have already done many times, to permute parameters. The most general
recursion obtained this way is rather formal, and is best described graphically:

\proclaim Theorem 6. Suppose that $z_j=q^2 z_i$. Then 
\eqn\thmfive{
\Psi_n\qquad\qquad\vbox{\hbox{\raise-1.9cm\llap{\epsfxsize=3.5cm\epsfbox{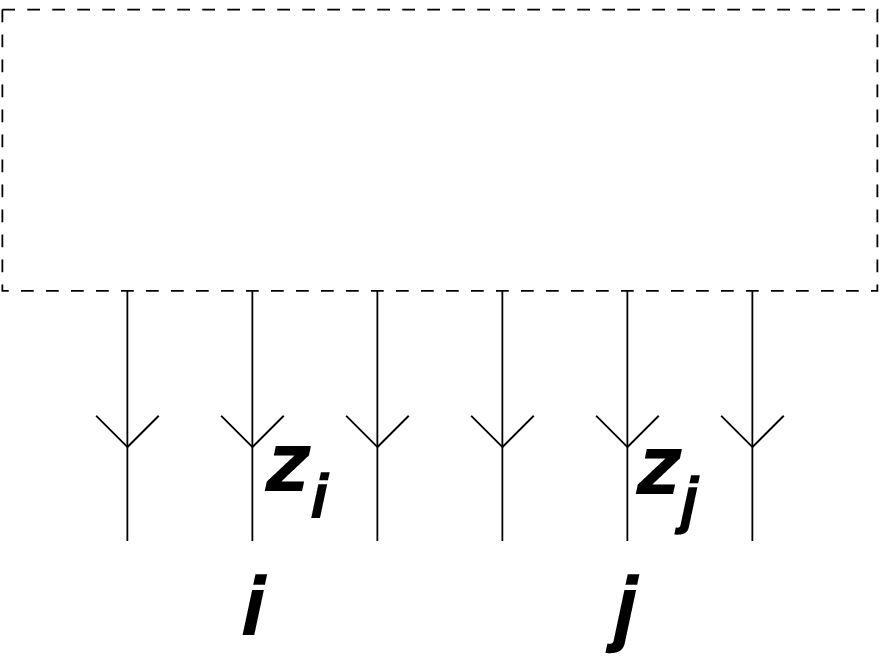}}}}
|_{z_j=q^2 z_i}=\left(\prod_{j<k<i} (q z_i-z_k)\right) 
\qquad\qquad\Psi_{n-1}\qquad\qquad\vbox{\hbox{\raise-1.9cm\llap{\epsfxsize=3.5cm\epsfbox{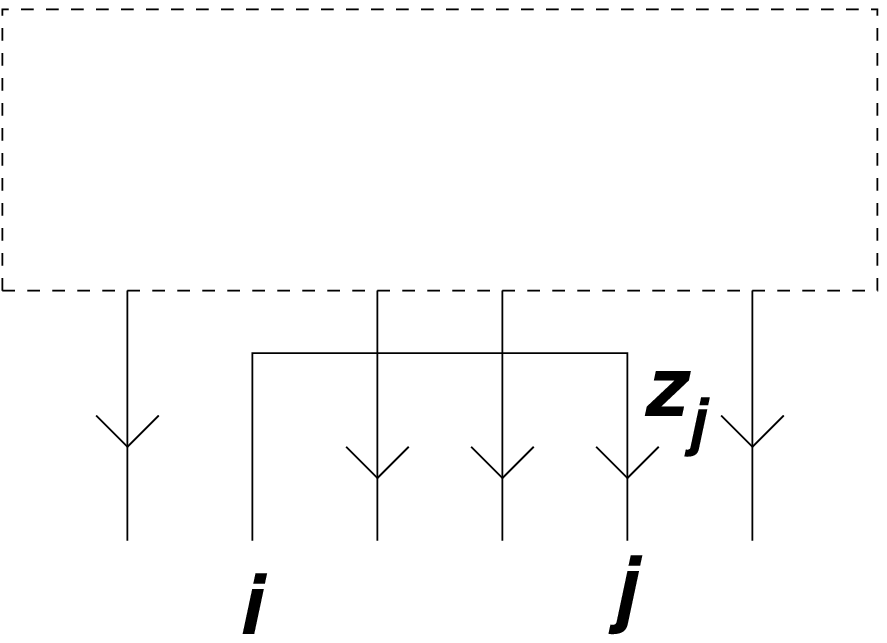}}}}
}

(cyclic order is implied in the range of the product).
Recall that each crossing represents an $R$-matrix. We have oriented the arch from $i$ to $j$
and attached to it the spectral parameter $z_j$, but we could have equally well oriented it from $j$ to $i$ and
given it the spectral parameter $z_i$, due to Eq.~\inv\ (up to modifying
the prefactor by a numerical constant).
The proof is elementary and proceeds graphically, using all three properties of Eqs.~\ybe--\inv.
Fixing the prefactor, which was the hard part in the proof of the Theorem 3, can now be simply
obtained by summing over all components and recovering the recursion of $Z_n$.
\qed

Among the consequences of Theorem 6, we obtain the property (iii) of section (3.2). Indeed to compute
the coefficient of the monomial $\prod_{k=1}^n (z_{i_k} z_{j_k})^{k-1}$ in $\Psi_n$, 
where $\{ \{i_1,\ldots,i_n\},\{j_1,\ldots,j_n\}\}$ is a (non-necessarily planar)
partition of $\{1,\ldots,2n\}$, it is sufficient to set
$z_{j_k}=q^2 z_{i_k}$, for all $k=1,2,\ldots,n$ and look for
the monomial $\prod_{k=1}^n z_{i_k}^{2(k-1)}$, as the partial degree property forbids distinct 
degrees for $z_{i_k}$ and $z_{j_k}$.
Applying iteratively Theorem 6 leads us to the evaluation of a certain diagram naturally associated
to the $i_k$ and $j_k$.
If we further assume that $i_k=\pi(j_k)$, $k=1,2,\ldots,n$ for some (planar) link pattern $\pi$,
then the diagram can be
transformed by use of Yang--Baxter and unitarity equations to the link pattern $\pi$, and
the weight of the monomial is easily computed to be $1$, thus proving the property.

A final general remark is in order, in view of the various recursion relations obtained here.
Theorem 1 shows that the complexity of the entry $\Psi_{n,\pi}$ as a polynomial
of the $z_i$ grows with the number
of little arches contained in the link pattern $\pi$. Indeed, the presence of few little arches
in the link patterns allows 
to factor out many terms from $\Psi_{n,\pi}$, corresponding to all sequences of points not separated
by little arches, thus lowering the degree of the remaining polynomial factor to be determined.
The latter is further constrained by Theorem 3 which proves sufficiently powerful to completely
fix $\Psi_{n,\pi}$ in the cases with small numbers ($2$, $3$, $4$) of little arches.
This is to be compared with the
results of \refs{\DFZJZ,\Kratt,\DFZ,\DFZJZb}, where the counting of FPL configurations was obtained up
to $4$ little arches.
The application of recursion relations (Theorems 3 and 6) to the actual computation
of components will be described in more detail in future work.

\subsec{Applications}

An immediate corollary of Theorem 5 obtained by taking the homogeneous limit where all the $z_j=1$
proves the conjecture that concerns the sum of
entries of $\Psi_n$ \BdGN, namely that
$3^{-n(n-1)/2} Z_n(1,1,\ldots,1)=A_n$ is the sum of all entries of the suitably normalized groundstate
of the homogeneous $O(1)$ loop model Hamiltonian. As mentioned before, $\Psi_n$ is
normalized by $\Psi_{n,\pi_0}=3^{n(n-1)/2}$, and therefore coincides with the groundstate vector
of the Hamiltonian $H_n=\sum_{i=1}^{2n} (I-e_i)$ up to the factor $3^{n(n-1)/2}$.
The appearance of the Hamiltonian may be seen for instance
by expanding the transfer matrix with all $z_j=1$ around $t=1$, 
so that at order $1$
in $(t-1)$ the eigenvector equation \main\ reduces to $H_n\Psi_n=0$.

Another corollary of Theorem 5 concerns the sum rules $P_n(k,t)$ of \PDFtwo\ for the case 
of the $O(1)$ loop model with $k$ dislocations on the boundary of the semi-infinite cylinder.
Indeed, the transfer matrices of \PDFtwo\ are obtained by simple restrictions of the
parameters of the general transfer
matrix $T'$ of Fig.~\zigzag, that commutes with our matrix $T$. 
To get the corresponding groundstate vectors, we simply have to set $2n-k$ parameters $z$ to $1$,
while the remaining $k$ all take the value $(1+q t)/(q+t)$. The sum rules are identified up to some
simple factor to the corresponding value of the IK determinant. The same holds for the two-parameter
refinement, leading to the sum rule \refasm. More generally, the polynomiality properties
observed in \PDFtwo\ can be inferred from those of the present work.

\newsec{Conclusion}

In this paper we have extended and proved a multi-parameter inhomogeneous version
of the sum rule \BdGN\ in the periodic case. On the way, we have been
able to derive recursion relations between the components of the groundstate vector,
and these might prove useful in understanding how the full Razumov--Stroganov conjecture,
concerning the individual entries of $\Psi_n$, should be generalized and hopefully proved.
Note however that the refined Razumov--Stroganov conjecture made in 
\PDFone\ in the case of the one-dislocation $O(1)$ loop model with one bulk parameter $t$
already involves partial summations of the entries of $\Psi_n(t)$ of the form
$\sum_\ell \Psi_{n,r^\ell \pi}(t)$, where $r\pi$ is the cyclically rotated version of
$\pi$ by one unit. These are necessary to ensure the cyclic covariance of these partial sums,
eventually identified with the corresponding sums of partition functions in the 6V DWBC model,
that connect the external bonds according to $\pi$ or any of its cyclically rotated versions.
This shows in the simplest case that multi-parameter generalizations 
of the full Razumov--Stroganov conjecture, if any exist, must be subtle.

The line of proof followed here should be applicable to other types of boundary conditions,
in relation to the versions of the 6V model with DWBC corresponding to other symmetry classes
of ASM, namely with the square grid possibly reduced to a smaller fundamental domain, with
accordingly modified boundary conditions. Indeed determinant or pfaffian formulae also exist
in these cases \refs{\KUPb,\OKADA}.

Another model of interest is the crossing $O(1)$ loop model, whose Hamiltonian on a 
semi-infinite cylinder of perimeter $n$
is expressed in terms of generators of the Brauer algebra \BRAU, and for which some entries 
of the groundstate vector were identified with degrees of algebraic varieties 
including the commuting variety. Preliminary
investigations show that simple inhomogeneous (one-parameter) generalizations of the model
produce a degree $2n-2$ groundstate vector with non-negative integer vector coefficients,
and suggest the existence of multi-parameter generalizations
with a nice polynomial structure and recursion relations
extending those of the present paper for the entries of the groundstate vector.
This will be pursued elsewhere. 
\bigskip

\noindent{\bf Acknowldegments}
We thank M.~Bauer, D.~Bernard, V.~Pasquier, Y.~Stroganov  
for discussions, and J.-B.~Zuber for a thorough reading of the manuscript.

\appendix{A}{The vector $\Psi_n(z_1,\ldots,z_{2n})$ for $n=2,3$}
We give below the explicit expressions for the vector $\Psi_n$ for $n=1,2,3$
as obtained directly by solving the eigenvector equation \main. For $n=1$,
the vector has a unique component, equal to $1$. For $n=2$, we find
\eqna\psitwo
$$
\eqalignno{ 
\Psi_{2,}\vcenter{\hbox{\epsfxsize=1.9cm\epsfbox{arch2-1.eps}}}
&
=q^2(q\,z_2-z_1)(z_3-q\,z_4) 
&\psitwo{a}
\cr
\qquad
\Psi_{2,}\vcenter{\hbox{\epsfxsize=1.9cm\epsfbox{arch2-2.eps}}}
&
=q^2(q\,z_3-z_2)(z_4-q\,z_1) 
&\psitwo{b}
\cr}$$
The normalization is such that $\Psi_2=3(1,1)$ when all $z_i=1$.
\goodbreak
For $n=3$, we have
\eqna\psithree
$$\openup\jot\eqalignno{
\Psi_{3,}\vcenter{\hbox{\epsfxsize=1.9cm\epsfbox{arch3-1.eps}}}
&=(q\,z_2-z_1)(q\,z_3-z_2)(q\,z_3-z_1)(z_4-q\,z_5)(z_5-q\,z_6)(z_4-q\,z_6)
\qquad&\psithree{a}\cr
\Psi_{3,}\vcenter{\hbox{\epsfxsize=1.9cm\epsfbox{arch3-4.eps}}}
&=(q\,z_4-z_3)(q\,z_5-z_4)(q\,z_5-z_3)(z_6-q\,z_1)(z_1-q\,z_2)(z_6-q\,z_2)
\qquad&\psithree{b}\cr
\Psi_{3,}\vcenter{\hbox{\epsfxsize=1.9cm\epsfbox{arch3-3.eps}}}
&=(q\,z_3-z_2)(q\,z_4-z_3)(q\,z_4-z_2)(z_5-q\,z_6)(z_6-q\,z_1)(z_5-q\,z_1)
\qquad&\psithree{c}\cr
\Psi_{3,}\vcenter{\hbox{\epsfxsize=1.9cm\epsfbox{arch3-5.eps}}}
&=(z_2-q\,z_3)(z_4-q\,z_5)(q\,z_1-z_6)\times \cr
\qquad&\!\!\!\!\!\!\!\!\!\!\!
\times \big((z_1-q\,z_2)(z_3-q\,z_4)(z_5-q\,z_6)+(z_4-q\,z_1)(z_2-q\,z_5)(z_6-q\,z_3)\big)
&\psithree{d}\cr
\Psi_{3,}\vcenter{\hbox{\epsfxsize=1.9cm\epsfbox{arch3-2.eps}}}
&=(z_3-q\,z_4)(z_5-q\,z_6)(q\,z_2-z_1)\times \cr
\qquad&\!\!\!\!\!\!\!\!\!\!\!
\times \big((z_2-q\,z_3)(z_4-q\,z_5)(z_6-q\,z_1)+(z_5-q\,z_2)(z_3-q\,z_6)(z_1-q\,z_4)\big)
&\psithree{e}\cr}$$
The normalization is such that $\Psi_3=27(1,1,1,2,2)$ when all $z_i=1$.

The reader will easily check 
Theorem 3 on 
\psitwo{} and \psithree{}.
For illustration of property (ii), at $n=3$, if we set $z_6=q^2 z_5$
and strip the link pattern $\vcenter{\hbox{\epsfxsize=1.1cm\epsfbox{arch3-5.eps}}}$ from its
little arch $5,6$, 
\eqnn\degethree{
$$\openup-10pt
\eqalignno{
\Psi_{3,}\vcenter{\hbox{\epsfxsize=1.9cm\epsfbox{arch3-3.eps}}}
\vert_{z_6=q^2 z_5}&= (z_2-q\,z_3)(z_4-q\,z_5)(q\,z_1-q^2z_5)(z_4-q\,z_1)(z_2-q\,z_5)(q^2z_5-q\,z_3)\cr
=&(q z_5-z_1)(q z_5-z_2)(q z_5-z_3)(q z_5-z_4)
\Psi_{2,}\vcenter{\hbox{\epsfxsize=1.9cm\epsfbox{arch2-2.eps}}}
&\degethree\cr}$$
which is nothing but Eq.~\recure\ for $i=5$.

\appendix{B}{Algebraic Bethe Ansatz and upper bound on the degree of $\Psi_n$}
In this appendix we construct the eigenvector $P_n$ using the Algebraic Bethe Ansatz.
As a corollary, we show that with a proper normalization, its components $\Psi_{n,\pi}$ are
polynomials of the inhomogeneities $z_i$ of total degree less or equal to $2 n^2(n+1)$.

To introduce the Algebraic Bethe Ansatz, we need to recall how the Temperley--Lieb loop
model can be recast in the framework of the six-vertex model. In much the same way as
the Temperley--Lieb loop Hamiltonian is equivalent to the twisted XXZ spin chain Hamiltonian in
a particular sector, (see e.g.\ \MNdGB), 
here our inhomogeneous transfer matrix is equivalent to the twisted inhomogeneous six-vertex
transfer matrix acting on the very same sector. We now introduce these objects.

The ``physical space'' of the six-vertex model consists
of $2n$ copies of ${\Bbb C}^2$; the auxiliary space is also ${\Bbb C}^2$.
The matrix $R_{i,0}=R(z_i,t)$ acts on the tensor product of the $i^{\rm th}$ space and the auxiliary space,
and is given by: (in the so-called homogeneous gradation)
\eqn\appR{
R(z,t)=\pmatrix{q\,z-q^{-1}t &0&0&0\cr 0&z-t&(q-q^{-1})t&0\cr 0&(q-q^{-1})z&z-t&0\cr 0&0&0&q\,z-q^{-1}t}
}

The monodromy matrix is
\eqn\appmono{
M_n(t|z_1,\ldots,z_{2n})=
R_{2n,0}(z_{2n},t)
\cdots R_{1,0}(z_1,t)
 }
It can be thought of as a $2\times 2$ matrix of operators acting on the physical space:
\eqn\appmonob{
M_n(t)=\pmatrix{A_n(t)&B_n(t)\cr C_n(t)&D_n(t)\cr}
}
The transfer matrix is the trace of the monodromy matrix
over the auxiliary space, but with a special twist:
\eqn\apptm{
T_n(t)=-q\, A_n(t)-q^{-1} D_n(t)
}

Consider then the following embedding of the space of link patterns into $({\Bbb C}^2)^{\otimes 2n}$.
To each $\pi\in LP_n$ we associate a vector obtained by taking the tensor product over
the set of arches of $\pi$, of the vectors $q^{1/2} \left({1\atop 0}\right)_j
\otimes\left({0\atop 1}\right)_k
-
q^{-1/2}\left({0\atop 1}\right)_j
\otimes\left({1\atop 0}\right)_k
$, where the indices $j<k$ are the endpoints of the arch, and
indicate the numbers of the pair of spaces $\Bbb C^2$ in which these vectors live.
Noting that 
\eqn\apptransl{
\check R= R{\cal P}=
(q\,z-q^{-1}t)I+(z-t)e\qquad e\equiv \pmatrix{0&0&0&0\cr 0&-q&1&0\cr 0&1&-q^{-1}&0\cr 0&0&0&0\cr}
}
and identifying $e$ with the usual Temperley--Lieb generator, we see that the $R$-matrix reproduces
the $R$-matrix introduced in the text (cf Eq.~\rrcop).
It can then be easily shown that $T_n$ leaves the subspace generated
by the $\pi$ stable, and that via the embedding
above its restriction is exactly our transfer matrix \transmat.

The Algebraic Bethe Ansatz is the following Ansatz for eigenstates of $T$:
\eqn\appABA{
P_n=\prod_{i=1}^{k} B_n(t_i)\cdot \left({1\atop 0}\right)^{\otimes 2n}
}
where the $t_i$ are some complex parameters
to be determined. Note that the matrices $B_n(t)$ commute for distinct values of $t$.
In the present situation, we set $k=n$.

A classical calculation (commutation of $B$'s with $A$'s and $D$'s)
shows that a sufficient condition for $P_n$ to be an eigenvector
of $T_n(t)$ is that the $t_i$ satisfy the Bethe Ansatz Equations (BAE). They can be recovered by
writing
the corresponding eigenvalue ${\cal T}_n(t)$ of the transfer matrix:
\eqn\appeig{
{\cal T}_n(t)\prod_{i=1}^k (t-t_i)=
-q\prod_{i=1}^{2n}(q^{-1} t-q\,z_i) \prod_{i=1}^k (q\,t-q^{-1} t_i)
-q^{-1}\prod_{i=1}^{2n}(t-z_i) \prod_{i=1}^k (q^{-1}t-q\,t_i)
}
and setting $t=t_i$ in it. Note that given ${\cal T}_n(t)$, 
Eq.~\appeig\ can be considered as a functional equation for the
function $Q_n(t)\equiv \prod_{i=1}^k (t-t_i)$ (so-called  $T$--$Q$ equation).

We now set $q=\e{2 i \pi/3}$, $k=n$ and
seek a function $Q_n(t)$ which satisfies Eq.~\appeig\ for which the eigenvalue has the form
${\cal T}_n(t)=\prod_{i=1}^{2n}(q\,t-q^{-1}z_i)$. Following Stroganov \STROIK, we notice that if we introduce
the function
\eqn\appF{
F_n(t)=\prod_{i=1}^{2n} (t-q\, z_i)\ \prod_{i=1}^n (t-t_i)
}
then one can rewrite Eq.~\appeig\ under the form:
\eqn\appFb{
F_n(t)+q^2 F_n(q\, t)+q\, F_n(q^2 t)=0
}
for all $t$. Since $F_n(t)$ is a polynomial of degree $3n$, one can expand it in powers of $t$
and one finds that Eq.~\appFb\ is equivalent to
\eqn\appFc{
F_n(t)=\sum_{i=0}^{3n} a_i t^i\quad\Rightarrow\quad a_{3k+1}=0 \qquad k=0,\ldots,n-1
}

Only remain $2n$ unknown coefficients $a_i$ ($a_{3n}=1$ by normalization), which are fixed by requiring
that $w_i\equiv q\, z_i$ be roots of $F_n(t)$. This leads to a system of linear equations for the $a_i$, which is
readily solved. One finds
\eqn\appFd{
F_n(t)={
\det\pmatrix{
1\hfill&\cdots\ &1\hfill&1\hfill\cr
w_1^2\hfill&\cdots\ &w_{2n}^2\hfill&t^2\hfill\cr
\vdots\hfill&&\vdots\hfill&\vdots\hfill\cr
w_1^{3k}\hfill&\cdots\ &w_{2n}^{3k}\hfill&t^{3k}\hfill\cr
w_1^{3k+2}\hfill&\cdots\ &w_{2n}^{3k+2}\hfill&t^{3k+2}\hfill\cr
\vdots\hfill&&\vdots\hfill&\vdots\hfill\cr
w_1^{3n}\hfill&\cdots\ &w_{2n}^{3n}\hfill&t^{3n}\hfill\cr
}
\over
\det\pmatrix{
1\hfill&\cdots\ &1\hfill\cr
w_1^2\hfill&\cdots\ &w_{2n}^2\hfill\cr
\vdots\hfill&&\vdots\hfill\cr
w_1^{3k}\hfill&\cdots\ &w_{2n}^{3k}\hfill\cr
w_1^{3k+2}\hfill&\cdots\ &w_{2n}^{3k+2}\hfill\cr
\vdots\hfill&&\vdots\hfill\cr
w_1^{3n-1}\hfill&\cdots\ &w_{2n}^{3n-1}\hfill\cr
}
}
}

We can identify these determinants with numerators in the Weyl formula for $GL(N)$ characters. More explicitly,
$Q_n$ is a ratio of two Schur functions:
\eqn\appQ{
Q_n(t)={
s_{Y_{n+1}}(w_1,\ldots,w_{2n},t)
\over
s_{\tilde{Y}_n}(w_1,\ldots,w_{2n})
}}
where $Y_{n+1}$ is the already introduced Young diagram with two rows of length $n$, two rows
of length $n-1$, $\ldots$, two rows of length $1$, and $\tilde{Y}_n$ is $Y_n$ with an extra row
of length $n$ added.

It is easy to check that ${\cal T}_n$ is generically a simple eigenvalue,
so that the vector $P_n$ we have just constructed (for $s_{\tilde Y_n}\ne 0$, which is also
generically true) must belong to the subspace of arches and identify via the
embedding above to the eigenvector $P_n$ of Eq.~\evectP\ (up to multiplication by a scalar). 
Let us now examine the dependence of the coefficients of $P_n$ as (rational) functions of the $z_i$.
Noting that the coefficients of the change of basis from the arches to the spin up/spin down
are constants and in particular independent of the
$z_i$, we define the degree of a vector-valued polynomial (in any given set of variables)
to be the (maximum) degree of its components in either basis.
We start from Eq.~\appABA. 
Each operator $B_n(t)$ is homogeneous of total degree $2n$ in all variables $z_i$ and $t$.
Therefore, as a function of the $z_i$ and of the $t_i$, $P_n$ is homogeneous of total degree $2n^2$, 
and is of partial degree $2n$ in each $t_i$. Furthermore it is a symmetric function of the $t_i$ by
construction, due to commutation of the $B_n(t_i)$. Therefore, it can be formally written as
\eqn\appdeg{
P_n=\sum_{\lambda, |\lambda|\le 2n^2, \lambda_1\le 2n}
p_{\lambda}(z_1,\ldots z_{2n}) s_\lambda(t_1,\ldots,t_n)
}
where $|\lambda|$ denotes the number of boxes of the Young diagram $\lambda$,
$\lambda_1$ is the length of its first row,
and $p_\lambda$ is some vector-valued
homogeneous polynomial in the $z_i$, of total degree $2n^2-|\lambda|$.

Now we assume that the $t_i$ are given by Eq.~\appQ, so that $P_n$ is the eigenvector of interest.
We can build the $s_\lambda(t_1,\ldots,t_n)$ out of the elementary symmetric functions 
$e_k(t_1,\ldots,t_n)$,
whose generating function is precisely $Q_n(t)=\prod_{i=1}^n (t-t_i)=\sum_{k=0}^n t^{n-k} (-1)^k e_k(t)$.
Each $s_\lambda$ is 
a sum of products of no more than $\lambda_1$ $e_{k_i}$ with
$\sum k_i=|\lambda|$ (indeed, the so-called Giambelli identity expresses 
the Schur function as a determinant: 
$s_\lambda=\det(e_{\lambda_i'-i+j})_{1\leq i,j\leq \lambda_1}$, where 
the $\lambda'_i$ are the lengths of columns of $\lambda$).
As $\lambda_1\leq 2n$, this means that
\eqn\appdegb{
s_\lambda(t_1,\ldots,t_n)={q_\lambda(w_1,\ldots,w_{2n})\over s_{\tilde{Y}_n}(w_1,\ldots,w_{2n})^{2n}}}
for some homogeneous symmetric polynomial $q_\lambda$ of total degree $2n^3+|\lambda|$.

Finally, combining Eqs.~\appdeg\ and \appdegb\ and substituting back $w_i=q\, z_i$, we find that 
$s_{Y_n}(z_1,\ldots,z_{2n})^{2n} P_n$ is a homogeneous polynomial of the $z_i$, of total
degree $2 n^2(n+1)$. $\Psi_n$ must divide it, hence the announced upper bound on the degree of $\Psi_n$.

\listrefs

\end